\documentclass[aps,prl,reprint,superscriptaddress,nofootinbib,nobibnotes]{revtex4-2}

\usepackage[T1]{fontenc}
\usepackage{lmodern}
\usepackage{amsmath,amssymb,bm,mathtools}
\usepackage{graphicx}
\usepackage{tikz}
\usepackage{bibunits}
\usepackage[colorlinks=true,citecolor=blue,linkcolor=black,urlcolor=blue]{hyperref}

\newcommand{\Tr}{\operatorname{Tr}}
\newcommand{\cZ}{\mathcal Z}
\newcommand{\HH}{\mathbb H}
\newcommand{\RR}{\mathbb R}
\newcommand{\ord}{\mathcal O}
\newcommand{\loc}{\mathrm{loc}}
\newcommand{\rot}{\mathrm{rot}}
\newcommand{\pert}{\mathrm{pert}}

\newcommand{\suppheading}[1]{\par\vspace{1.25em}\begin{center}\textbf{#1}\end{center}\vspace{-0.40em}\par}
\newcommand{\suppsubheading}[1]{\par\vspace{0.85em}\begin{center}\textbf{#1}\end{center}\vspace{-0.45em}\par}

\newcommand{\EH}{\mathrm{EH}}
\newcommand{\GH}{\mathrm{GH}}
\newcommand{\ct}{\mathrm{ct}}

\newcommand{\ESU}{\mathrm{ESU}}

\newcommand{\scal}{\mathrm{sc}}
\newcommand{\Dir}{\mathrm D}

\newcounter{prlappendix}

\begin{document}

\title{Rotating the Replica: Twisted R\'enyi Coefficients from CFT Operator Counting}

\author{Jyotirmoy Mukherjee}
\email{jyotirmoy.mukherjee@kcl.ac.uk}
\affiliation{Department of Mathematics,
King's College London,
Strand, London WC2R 2LS, United Kingdom}
\begin{abstract}
We show that twisting vacuum R\'enyi entropies by a spatial rotation reveals conformal field theory data governing the asymptotic counting of spinning operators. For a ball in four dimensions, the logarithmic
coefficients of these angular-momentum-twisted R\'enyi entropies are
determined by rotating-sphere thermal data to all orders in the local
derivative expansion, including the Weyl-anomaly contribution. We test
this relation in free theories and in strongly coupled, large-$N$
$\mathcal N=4$ Yang--Mills theory using holography.
\end{abstract}
\maketitle


\begin{bibunit}[apsrev4-2]

\section{Introduction}

Conformal field theories (CFTs) are important in the study of critical phenomena and quantum gravity, particularly through the AdS/CFT correspondence~\cite{Maldacena:1997re}.
Two important aspects of a CFT are its spectrum of local operators and the entanglement structure of its vacuum. The former
characterizes the spectrum of states on the sphere through the
state--operator correspondence, whereas the latter
probes how the degrees of freedom are entangled in a spatial subregion. Whether information about one can be
extracted from the other is not immediately clear.
It has been shown that the high-temperature partition function of
rotating CFTs is systematically organized by a local
thermal effective field theory (EFT)
~\cite{BhattacharyyaRotating,BanerjeeEquilibrium,
JensenEquilibrium}. More recently, this framework
has been used to study the asymptotic spectrum of
CFT operators, including large-spin limits and
subleading thermal corrections
~\cite{BenjaminOoguri,AnandSemiUniversality,
KomargodskiRegge,FromZtoa}.
For more recent developments along
these lines, see
\cite{Kim:2023sig, Bajaj:2024utv,Lee:2026azy,Mukherjee:2026dfu,Mishra:2026bpi,
David:2026yis,Miscioscia:2026bvl}.

For a spherical region, the Casini--Huerta--Myers (CHM) map relates the vacuum reduced density matrix to the thermal density matrix on the  Euclidean geometry $S^1\times\HH^{d-1}$~\cite{CHM}.
Using this conformal map, the vacuum entanglement and
R\'enyi entropies across spherical entangling surfaces
have been computed from the thermal partition function
on the hyperbolic cylinder in a variety of theories,
including free fields, gauge and higher-spin theories,
and holographic CFTs
~\cite{CasiniHuertaSphere,KlebanovFreeRenyi,
HungMyersSmolkinYale,DavidMukherjee2021,
MukherjeeHigherDerivative,MukherjeeHigherSpin}.
More recently, it has been combined with the thermal EFT to show that
the leading small-$n$ behavior of the untwisted spherical R\'enyi
entropy is universal and fixed by the equation of state 
~\cite{KusukiOoguriPal}.

These developments raise a natural question: can universal information
about entanglement be extracted from the high-temperature data that
govern the asymptotic operator spectrum? We show that this is the case
for R\'enyi entropies twisted by a spatial rotation. In the local thermal-EFT sector, the twist produces a leading small-\(n\) logarithmic contribution whose coefficients probe higher-derivative rotating-sphere thermal data, beyond the equation of state.

\emph{Angular-momentum-twisted R\'enyi partition function}--
We consider a four-dimensional CFT in its vacuum state, defined on
flat spacetime with Euclidean metric
\begin{equation}\label{flatmet}
    ds^2=dt_E^2+dr^2+r^2d\Omega_2^2.
\end{equation}
We choose a ball-shaped spatial subregion $A$ of radius $R$,
whose entangling surface is the two-sphere $\partial A=S^2$
at $t_E=0$ and $r=R$. The corresponding vacuum reduced density
matrix is
$\rho_A=\Tr_{\bar A}|0\rangle\langle 0|$. Here $\bar{A}$ denotes the region outside of the ball-shaped region.

We wish to probe the angular-momentum dependence of the entanglement spectrum  by introducing the rotation restricted to $A$,
$U_\theta=e^{i\theta J_A}$, about an axis passing through the
center of the ball, where $J_A$ is the corresponding angular-momentum
generator. Since both the vacuum and the spherical subregion
are invariant under this rotation, the reduced density matrix
commutes with $J_A$, namely $[\rho_A,J_A]=0$.
Consequently, $\rho_A$ and $J_A$ admit a common eigenbasis,
\begin{equation}
    \rho_A|\lambda,j\rangle=\lambda|\lambda,j\rangle,
    \qquad
    J_A|\lambda,j\rangle=j|\lambda,j\rangle.
\end{equation}
We therefore define the angular-momentum-twisted R\'enyi
partition function as
\begin{equation}
    \cZ_n(\theta)
    \equiv \Tr\left(\rho_A^n U_\theta\right)
    =\sum_{\lambda,j}\lambda^n e^{i\theta j}.
\end{equation}
The eigenvalues of the reduced density matrix
are thus weighted by phases determined by their angular
momentum quantum numbers. In this way, the twisted partition
function encodes the angular momentum dependence of the
R\'enyi partition function.

Since we are considering the vacuum reduced density matrix
of a ball-shaped region, we can employ the conformal map
of Casini, Huerta, and Myers (CHM)~\cite{CHM}, which maps
the causal development of the ball to the hyperbolic
cylinder $S^1\times\mathbb H^3$.
In Euclidean signature, the conformal transformation is given by
\begin{equation}
    t_E=\frac{R\sin\tau}{\cosh u+\cos\tau},
    \qquad
    r=\frac{R\sinh u}{\cosh u+\cos\tau}.
\end{equation}
Under this transformation, the flat metric \eqref{flatmet}
takes the form
\begin{equation}
\begin{aligned}\label{hypmet}
    ds^2_{\rm flat}
    &=\Omega^{-2}\left(
    d\tau^2+du^2+\sinh^2u\,d\Omega_2^2
    \right)\\
    &=\Omega^{-2}ds^2_{S^1\times\mathbb H^3},
\end{aligned}
\end{equation}
where the conformal factor is
\begin{equation}
    \Omega=\frac{\cosh u+\cos\tau}{R}.
\end{equation}

The conformal transformation leaves the angular
coordinates $(\vartheta,\phi)$ of $S^2$ unchanged.
Consequently, $J_A$ maps to the spatial rotation
generator $J=-i\partial_\phi$ on $\mathbb H^3$.
Since the reduced density matrix is mapped to a
thermal density matrix on $\mathbb H^3$, the twisted
R\'enyi partition function becomes
\begin{equation}
\begin{aligned}
\cZ_n(\theta)
&=\frac{\Tr_{\mathbb H^3}
\left(e^{-2\pi nH}U_\theta\right)}
{\left[\Tr_{\mathbb H^3}
\left(e^{-2\pi H}\right)\right]^n},
\end{aligned}
\label{eq:twisted}
\end{equation}
where $H$ generates hyperbolic time translations. The insertion of $U_\theta$ is implemented by the twisted
identification of $(\tau,\phi)\sim(\tau+2\pi n,\phi+\theta)$.  In the
co-rotating angle $\phi-\theta\tau/2\pi n$, it becomes an untwisted identification with an off-diagonal, rotating-frame metric.

The ordinary R\'enyi entropy and a normalized angular-momentum-twisted
R\'enyi generating functional are defined by, respectively,
\begin{equation}
\begin{aligned}
 S_n(A)&=\frac{1}{1-n}\log\cZ_n(0),\\
 S_n(\theta)&=\frac{1}{1-n}\log\!\left[
 \frac{\cZ_n(\theta)}{\cZ_1(\theta)^n}\right].
\end{aligned}
\label{eq:entropydef}
\end{equation}

\section{Principal results}

\emph{Local logarithmic R\'enyi coefficients.}---
We find that, in the local thermal-EFT regime
$2\pi n\ll|\theta|\ll1$, the logarithmic contribution
of the twisted R\'enyi entropy is governed
by high-temperature CFT data. In the leading small-\(n\) limit, the local contribution takes the  following form
\begin{equation}
\left.\lim_{n\to0}nS_n(\theta)\right|_{\loc,\pert}
=\left[\frac{c'}{\theta^2}
+2\sum_{q\ge1}s_q^{\rm phys,\loc}\theta^{2q-2}\right]
\log\frac{R}{\epsilon}.
\label{eq:leading}
\end{equation}
Here, $R$ and $\epsilon$ are the radius of the ball and the UV cutoff, respectively, and “loc, pert” denotes the local thermal-EFT contribution, extracted by first taking the leading small-\(n\) limit and then expanding in \(\theta\). The coefficients \(s_q^{\rm phys,loc}\) are obtained from the  partition function on \(S^1\times\mathbb H^3\), including its finite Weyl-anomaly contribution.
The coefficient $c'$ is determined by the flat-space
equation of state,
$\log Z_{\RR^3\times S^1_\beta}\sim c'V/\beta^3$.

Our main result is that these coefficients are fixed by the
high-temperature data of the rotating sphere,
\begin{equation}
s_q^{\rm phys,\loc}
=-\frac{a_{0,q;0}^{\rm trace}}{4\pi^2},
\qquad q\ge1.
\label{eq:physicalmapmt}
\end{equation}
Here $a_{m,k;p}^{\rm trace}$ are the coefficients of the
high-temperature expansion of the physical thermal trace on the
rotating $S^3$, including the Weyl-anomaly term~\cite{FromZtoa,Advant:2026zft}.
We first establish this relation for the Weyl-invariant local thermal actions and then show that the finite Wess–Zumino contributions agree under the same leading small-\(n\) projection.
Thus, the local logarithmic coefficients of twisted Rényi entropies are fixed by the same high-temperature CFT data that govern asymptotic operator counting, without requiring an independent entanglement calculation.

\section{Geometric origin of the R\'enyi logarithm}

To evaluate the thermal EFT on the twisted hyperbolic
cylinder, we introduce cylindrical coordinates adapted
to the rotational fixed axis:
\begin{equation}
\sinh r=\sinh u\sin\vartheta,\qquad
\tanh\eta=\tanh u\cos\vartheta.
\label{eq:coordtrans}
\end{equation}
The Euclidean hyperbolic metric in \eqref{hypmet}
and thermal identification vector become
\begin{equation}
\begin{aligned}
ds^2&=d\tau^2+dr^2+\cosh^2r\,d\eta^2
      +\sinh^2r\,d\phi^2,\\
v&=(2\pi n)\,\partial_\tau+\theta\,\partial_\phi,
\end{aligned}
\label{eq:hyperbolic}
\end{equation}

At leading order, the local thermal EFT is determined
entirely by the flat-space high-temperature coefficient
$c'$.
Since the proper length of the local thermal cycle is
$\sqrt{v^2}$, its leading Weyl-invariant contribution is
\begin{equation}
\log Z_{\loc}^{(0)}
=c'\int d^4x\,\frac{\sqrt g}{(v^2)^2}.
\label{eq:ideal}
\end{equation}
Since $\mathbb H^3$ is noncompact, the spatial integral
requires a regulator. For a nonzero angular twist,
the transverse integral is convergent, while the
integration along the rotational fixed geodesic
remains divergent. We regulate this divergence using
the UV cutoff $\epsilon$ of the original spherical
region. Under the CHM map, the corresponding hyperbolic
cutoff is
$u_{\max}=\log(2R/\epsilon)+\ord(\epsilon/R)$.
Since $u=|\eta|$ along the fixed geodesic $r=0$,
its regulated length is
\begin{equation}
L=\int_{-u_{\max}}^{u_{\max}}d\eta
=2\log\frac{R}{\epsilon}+\ord(1).
\label{eq:L}
\end{equation}
We can now evaluate Eq.~\eqref{eq:ideal}
by integrating over the thermal cycle, the angular
coordinate $\phi$, and the transverse radial direction.
At logarithmic order, we obtain
\begin{equation}
\begin{aligned}
\log Z_{\loc}^{(0)}
&\simeq (2\pi)(2\pi n)c'L
\int_0^\infty dr\,
\frac{\sinh r\cosh r}
{[(2\pi n)^2+\theta^2\sinh^2r]^2}\\
&=\frac{c'L}{2n\theta^2}.
\end{aligned}
\label{eq:idealresult}
\end{equation}
Since the Weyl-invariant integrand depends only on $r$, the inherited
CHM cutoff restricts the $\eta$-interval to length
$L-2\log\cosh r+\cdots$. At fixed $n$ and $\theta$, the second term
contributes only to the cutoff-independent part of the convergent
transverse integral. It therefore does not affect the coefficient
of $\log(R/\epsilon)$, allowing us to factor out $L$ from the local
partition function when extracting the logarithmic R\'enyi coefficient.
The radial integral is dominated by a thin tube $r\sim 2\pi n/|\theta|$ around the fixed axis, where the local thermal length $\sqrt{v^2}\sim 2\pi n$ is small and the thermal EFT is reliable; the EFT fails only at $|\theta|\sinh r\gtrsim 1$, a region whose contribution is subleading at small $n$. The twist thus replaces the $1/n^3$ volume scaling of the untwisted case~\cite{KusukiOoguriPal} by $c'(n/\theta)^2/n^3\propto 1/(n\theta^2)$, and at leading order in $n$ the divergence is logarithmic rather than area-law.
Thus, the logarithmic dependence on the UV cutoff
arises from the regulated length of the fixed geodesic,
while its coefficient is determined by the local
thermal EFT. Using Eq.~\eqref{eq:L}, we reproduce
the leading term of Eq.~\eqref{eq:leading}.

\section{All-order R\'enyi coefficient relations with thermal CFT data}

We now establish the relation between the logarithmic
R\'enyi coefficients and the high-temperature
operator-counting data of the CFT.
The key point here is that both $S^1\times S^3$ and
$S^1\times\HH^3$ are locally conformally flat.
Consequently, the Weyl-invariant local thermal effective actions on these backgrounds are built from the same geometric invariants, evaluated with different thermal identifications.

The correspondence can be understood geometrically.
The spherical entangling surface has two fixed points,
the north and south poles, under a spatial rotation
about the axis passing through them. A conformal transformation can map the entangling sphere
to a plane by sending either of its two poles to infinity.
In this description, we can identify two orthogonal planes
of rotation of $S^3$ (embedded in $\mathbb R^4$): one generates the replica
rotation, while the other generates the angular momentum
rotation that leaves the poles fixed.

The combined identification is therefore generated by
\begin{equation}
    v=2\pi n\,J_{\perp}+\theta\,J_{\parallel},
    \label{eq:geometricrot}
\end{equation}
with angular displacements $(2\pi n,\theta)$.
To make the geometric correspondence explicit,
consider the rotating-sphere cylinder with metric
\begin{equation}
ds_S^2=d\eta^2+d\psi^2
+\sin^2\psi\,d\tau^2
+\cos^2\psi\,d\phi^2,
\label{eq:geomSphereCylinder}
\end{equation}
where $\eta$ is the thermal coordinate and
$(\tau,\phi)$ are the two angular coordinates.
On the formal $\beta_S=0$ slice of the
rotating-sphere thermal EFT, the identification
vector is
$v_S=(2\pi n)\partial_\tau+\theta\partial_\phi$.

Under the coordinate transformation
\begin{equation}
\sin\psi=\frac{1}{\cosh r},
\qquad
\cos\psi=\tanh r,
\label{eq:geomCoordinateMap}
\end{equation}
the rotating-sphere and hyperbolic-cylinder
metric--thermal-vector pairs are locally related by
\begin{equation}
g_S=\frac{1}{\cosh^2 r}\,g_H,
\qquad
v_S=v_H.
\label{eq:geomWeylRelation}
\end{equation}
Consequently, the local Weyl-invariant thermal-EFT densities considered here agree on the two geometries, independently of derivative order.
The relative normalization of the integrated
actions follows from their different integration
ranges and periods, giving the factor $nL$,
where $L=2\log(R/\epsilon)+O(1)$.
The explicit matching of the thermal-EFT densities
and integration measures is given in Appendix~\ref{appB}.

It was shown in Refs.~\cite{FromZtoa}
that on conformally flat backgrounds, the local
Weyl-invariant thermal effective action can be
expressed in terms of a finite set of geometric
invariants. Since the twisted hyperbolic cylinder
is also conformally flat, and the thermal
identification is generated by a Killing vector $v^{\mu}$,
the same classification applies here.
In four dimensions, the relevant scalar invariants
are the vorticity scalar $\Tr\omega^2$,
the Weyl-covariant Schouten scalar $v^2\mathcal S$,
and the contraction
$v^2\mathcal B_\mu(\omega^2)^{\mu\nu}\mathcal B_\nu$,
where $\omega_{\mu\nu}$ is the vorticity tensor,
$\mathcal S_{\mu\nu}$ is the Weyl-covariant Schouten
tensor, and $\mathcal B_\mu=\mathcal S_{\mu\nu}v^\nu$.
We use this basis to evaluate the thermal effective
action on the hyperbolic cylinder and the rotating
sphere. The conventions and explicit calculations
are given in Appendix~\ref{appA}.

The local thermal action on $S^1\times \HH^3$ is
\begin{equation}
\log Z_{\loc}[g,v]
=\int d^4x\,\frac{\sqrt g}{(v^2)^2}
\left[c'+\sum_{j\ge1}\mathcal I_{2j}[g,v]\right],
\label{eq:generaleft}
\end{equation}
where $\mathcal I_{2j}$ denotes the Weyl-invariant local
terms at derivative order $2j$. Their coefficients
are intrinsic thermal data of the CFT and are
independent of the background on which the action
is evaluated.

Evaluating Eq.~\eqref{eq:generaleft} on the twisted
hyperbolic cylinder, we obtain
\begin{equation}
\begin{aligned}
\log Z_{\HH^3}^{\loc}
=2\pi(2\pi n)L
\int_0^\infty dr\,
\frac{\sinh r\cosh r\,\mathcal I_{\HH^3}(r)}
{[(2\pi n)^2+\theta^2\sinh^2r]^2},
\end{aligned}
\label{eq:hyperradial}
\end{equation}
where $\mathcal I_{\HH^3}$ denotes the complete
local derivative expansion, including $c'$, the leading coefficient.
Let us evaluate the
Weyl-invariant thermal action on the twisted hyperbolic
cylinder. Introducing the radial variable
\begin{equation}
z=\frac{(2\pi n)^2\theta^2\cosh^2r}
{(2\pi n)^2+\theta^2\sinh^2r},
\label{eq:zvar}
\end{equation}
we find that the three Weyl-invariant scalars
introduced above reduce to

\begin{equation}
\begin{aligned}
\frac18\Tr\omega^2&=-z,\\
v^2\mathcal S&=3z-(2\pi n)^2-\theta^2,\\
v^2\mathcal B_\mu(\omega^2)^{\mu\nu}\mathcal B_\nu
&=4z[z-(2\pi n)^2](z-\theta^2).
\end{aligned}
\label{eq:hyperinv}
\end{equation}

Moreover, the change of variables in
Eq.~\eqref{eq:zvar} gives the integral measure
\begin{equation}
\frac{\sinh r\cosh r\,dr}
{\left[(2\pi n)^2+\theta^2\sinh^2r\right]^2}
=
\frac{dz}
{2(2\pi n)^2\theta^2
[(2\pi n)^2-\theta^2]}.
\label{eq:measurematch}
\end{equation}
Therefore, the local thermal effective action on the
twisted hyperbolic cylinder takes the form
\begin{equation}
\begin{aligned}
\log Z_{\HH^3}^{\loc}
&=\frac{\pi(2\pi n)L}
{(2\pi n)^2\theta^2[(2\pi n)^2-\theta^2]}
\int_{\theta^2}^{(2\pi n)^2}
 dz\,\mathcal I_{\HH^3}(z).
\end{aligned}
\label{eq:hyperintegral}
\end{equation}
Since the Weyl-invariant scalars are polynomials
in $z$, $(2\pi n)^2$, and $\theta^2$, the integral
at each derivative order reduces to a homogeneous
polynomial in $(2\pi n)^2$ and $\theta^2$.
Consequently, the complete local derivative expansion
can be organized as
\begin{equation}
\log Z_{\HH^3}^{\loc}
=-\frac{L}{4\pi^2n\theta^2}
\sum_{m,k\ge0}\widehat a_{m,k}
(4\pi^2n^2)^m\theta^{2k},
\label{eq:hyperexpansion}
\end{equation}
where $\widehat a_{m,k}$ are combinations of the
theory-dependent thermal-EFT Wilson coefficients.

We now identify these combinations with the
high-temperature rotating-$S^3$ data.
The Weyl-invariant rotating-sphere partition function
admits the expansion~\cite{FromZtoa,Advant:2026zft}
\begin{equation}
\begin{aligned}
\frac{\log Z_{S^3}^{\rot}}{\beta}
&=-\frac{1}{y_1y_2}
\sum_{m,k,p\ge0}a_{m,k;p}^{\rot}
y_1^m y_2^k\beta^{2p},\\
y_i&=\beta^2+\alpha_i^2,
\end{aligned}
\label{eq:tracecoeff}
\end{equation}
where $\beta$ is the spherical thermal period and
$\alpha_{1,2}$ are angular displacements along its
two commuting rotations.

By explicitly matching the Weyl-invariant generators
and their reduced integration measures on the two
geometries, we find that under the formal continuation
\begin{equation}
\beta=0,\qquad
\alpha_1=2\pi n,\qquad
\alpha_2=\theta,
\label{eq:continuation}
\end{equation}
the coefficients obtained from the hyperbolic action
are precisely those of the rotating-sphere EFT:
\begin{equation}
\widehat a_{m,k}=a_{m,k;0}^{\rot}.
\label{eq:coefficientmatch}
\end{equation}
In other words, the local partition functions are related by
\begin{equation}
\log Z_{\HH^3}^{\loc}(n,\theta)
=nL\left.
\frac{\log Z_{S^3}^{\rot}(\beta,\alpha_1,\alpha_2)}
{\beta}
\right|_{\substack{\beta=0\\
\alpha_1=2\pi n,\;\alpha_2=\theta}}.
\label{eq:Gidentity}
\end{equation}
The matching holds at every order in the
Weyl-invariant local derivative expansion,
independently of the underlying CFT.
The explicit derivation, including the relative
normalization, is given in Appendix~\ref{appB}.

The leading small-$n$ contribution to
Eq.~\eqref{eq:hyperexpansion} comes from $m=0$.
Using $L=2\log(R/\epsilon)+\ord(1)$, we obtain
\begin{equation}
\left.\lim_{n\to0}nS_n(\theta)\right|_{\loc,\pert}^{\rm WI}
=-\frac{\log(R/\epsilon)}{2\pi^2}
\sum_{k\ge0}a_{0,k;0}^{\rot}\theta^{2k-2}.
\label{eq:renyimap}
\end{equation}
Equation \(\eqref{eq:renyimap}\) establishes the Weyl-invariant coefficient relation, with \(a_{0,0;0}^{\rm rot}=-2\pi^2c'\), at every order in the local thermal EFT.

\emph{Weyl anomaly and physical coefficients.}---
The relation \eqref{eq:renyimap} shows that the
angular-momentum-twisted R\'enyi entropy, expanded
first in the small replica index and then in the
small rotation angle, is directly related to the
high-temperature CFT data encoded in the rotating
sphere partition function.
However, the physical trace on the rotating-sphere  differs from
the Weyl-invariant result by a finite Weyl anomaly,
$\Delta_S=\log Z_{\rm trace}-\log Z_{\rot}$.
In the conventions of Ref.~\cite{FromZtoa},
\begin{equation}
\frac{\Delta_S}{\beta_S}
=\frac{a_{\rm W}}{12}
\left(\frac{y_1}{y_2}+\frac{y_2}{y_1}+7\right),
\qquad y_i=\beta_S^2+\alpha_i^2.
\label{eq:sphereanomaly}
\end{equation}
Similarly, on the hyperbolic cylinder,
the physical partition function differs from the
Weyl-invariant one by a Wess--Zumino contribution,
$\Delta_H=\log Z_H^{\rm phys}-\log Z_H^{\rm WI}$.
We work in the $b=0$ curvature-counterterm scheme
on both sides, including Eq.~\eqref{eq:sphereanomaly}.
Adding a finite $\int\sqrt g\,R^2$ counterterm
changes the split between the Weyl-invariant and
anomalous pieces (including $s_2^{\rm WI}$ in the
unit-thermal frame), but not their consistently
combined physical-frame result.
Although the two Weyl transformations are different,
we find that their contributions agree under the
same projection as in Eq.~\eqref{eq:Gidentity}.
In particular, in the leading small-$n$ limit,
\begin{align}
\left.\lim_{n\to0}\frac{n}{L}\Delta_H
\right|_{\loc,\pert}
&=
\lim_{n\to0}n^2
\left.\frac{\Delta_S}{\beta_S}\right|
_{\substack{\beta_S=0\\
\alpha_1=2\pi n,\;\alpha_2=\theta}}
\nonumber\\
&=\frac{a_{\rm W}\theta^2}{48\pi^2}.
\label{eq:hyperanomaly}
\end{align}
The hyperbolic Wess--Zumino contribution is
computed independently in Appendix~\ref{appC}.
Therefore, after including the anomaly on both
sides, the correspondence extends to the
physical coefficients in Eq.~\eqref{eq:physicalmapmt}.
This relation holds for the local perturbative
coefficients in the leading small-$n$ expansion.

\section{Tests of the R\'enyi coefficient relations}

\emph{Holographic test.}---
We first provide a strong-coupling check using
five-dimensional Einstein gravity dual to large-$N$
$\mathcal N=4$ Yang--Mills theory.
We evaluate the local action
of a Euclidean rotating topological AdS$_5$ solution
~\cite{Dehghani,HeRotating}, using standard holographic
renormalization~\cite{BalasubramanianKraus,Skenderis,HenningsonSkenderis}.
Writing $r_+$ for its horizon coordinate and $\alpha$
for the rotation parameter, its radial
function is
$\Delta(\hat r)=(\hat r^2-\alpha^2)(\hat r^2-1)-2m$.
Smoothness determines the physical cycles,
\begin{equation}
\begin{aligned}
 \beta_H&=\frac{2\pi(r_+^2-\alpha^2)}
 {r_+(2r_+^2-\alpha^2-1)},\\
 \theta&=-\frac{2\pi\alpha(r_+^2-1)}
 {r_+(2r_+^2-\alpha^2-1)},
\end{aligned}
\label{eq:holographiccycles}
\end{equation}
where $\beta_H=2\pi n$. We obtain the local renormalized gravitational action by
combining the on-shell Einstein--Hilbert action, the
Gibbons--Hawking--York boundary term, and the holographic
counterterms. Since the induced boundary metric is
conformally related to the physical hyperbolic cylinder,
\begin{equation}
g_b=D^{-1}g_H,\qquad
D=1+\alpha^2\sinh^2r,
\end{equation}
we must also include the corresponding finite
Wess--Zumino contribution to account for the Weyl
anomaly and the associated conformal-frame dependence
of the partition function.

In the leading small-$n$ sector, let
$K_S^{\pert}(\theta)$ denote the coefficient defined by
$\lim_{n\to0}nS_n|_{\loc,\pert}=L K_S^{\pert}(\theta)$.
Combining the renormalized gravitational on-shell action
with the finite Wess--Zumino contribution, we obtain
\begin{equation}
\begin{aligned}
K_{S,\mathcal N=4}^{\pert}(\theta)
&=N^2\frac{(1-q^2)^3}{4q^2(2-q^2)^2}\\
&=N^2\left[
\frac{\pi^2}{16\theta^2}-\frac1{16}
+\frac{\theta^2}{32\pi^2}
-\frac{\theta^4}{64\pi^4}
+\ord(\theta^6)\right],
\end{aligned}
\label{eq:strong}
\end{equation}
where $q=\alpha/r_+$ is related to the physical
angular twist by $\theta=-2\pi q/(2-q^2)$
in the leading small-$n$ limit.
The first line determines the complete leading
local high-temperature function, rather than
only its small-angle coefficients.

The Wess--Zumino contribution here relates the natural
black-hole boundary frame to $g_H$, rather than the
unit-thermal-length frame used in Appendix~\ref{appC}.
The distinct frame transformations explain its nonzero
$\theta^4$ term, as checked explicitly in that appendix.

We independently verify the coefficients using the
renormalized on-shell action of singly rotating
Kerr--AdS$_5$, with the conformal boundary chosen
to be the round $S^1\times S^3$
~\cite{GibbonsPerryPope,FromZtoa}. By symmetry under
exchange of the two rotation planes,
$a_{m,0;0}=a_{0,m;0}$, so the leading small-$n$
projection selects the one-spin series
$\sum_{m\ge0}a_{0,m;0}\theta^{2m-2}$.
Normalizing by the sphere's fixed-circle length
$2\pi$, rather than the hyperbolic length $L$,
gives agreement through $\ord(\theta^4)$.
The details are provided in the
Supplemental Material~\cite{Supplement}.

\emph{Free CFT checks.}---
We  verify the perturbative coefficient
relation for free scalar and Dirac fields
using hyperbolic heat kernels and spin characters.
We obtain
\begin{equation}
\begin{aligned}
K_{S,\rm sc}^{\pert}(\theta)
&=\frac{\pi^2}{180\theta^2}+\frac1{144}
-\frac{\theta^2}{960\pi^2},\\
K_{S,\rm D}^{\pert}(\theta)
&=\frac{7\pi^2}{360\theta^2}-\frac1{144}
-\frac{7\theta^2}{1920\pi^2},
\end{aligned}
\label{eq:free}
\end{equation}
The $\theta^2$ terms agree with the corresponding
$-a_{0,2;0}^{\rm trace}/(4\pi^2)$ from the physical
rotating-sphere traces, without adding the anomaly
shift a second time. It will be interesting to test this phenomenon in exotic free fields \cite{Mukherjee:2026dfu} including gauge theories where edge modes require separate
treatment~\cite{DonnellyWall,SoniTrivedi2016, DavidMukherjee}.

\section{Discussion}

In this Letter, we have related an information-theoretic
quantity, the angular-momentum-twisted R\'enyi entropy
across a spherical entangling surface, to the
high-temperature CFT data on $S^3$ that govern the
asymptotic counting of local operators.
The origin of this connection lies in the locality
of the thermal effective action: the same local
Weyl-invariant densities determine both quantities,
while their integration measures establish the
coefficient correspondence.
Thus, although the individual coefficients depend
on the underlying CFT, the relation between them
is universal. The leading R\'enyi coefficient is
determined by the high-temperature equation of state,
whereas the subleading coefficients probe
higher-derivative thermal data.
We have independently tested this relation in free
field theories and in strongly coupled, large-$N$
$\mathcal N=4$ Yang--Mills theory using a
local gravitational calculation in AdS$_5$.
The agreement provides further evidence for the
role of locality in organizing universal properties
of CFT entanglement.

An interesting extension is to include global
symmetries and study R\'enyi entropies twisted by
both angular momentum and conserved charges, with
connections to charged and symmetry-resolved
entanglement~\cite{BelinCharged,GoldsteinSela,BeniniAsymmetry}.
In this case, the thermal effective action also
depends on background gauge fields and chemical
potentials~\cite{BanerjeeEquilibrium,JensenEquilibrium}.
It would be interesting to determine how the
additional thermal data and possible  anomalies
enter the R\'enyi coefficients.

Another direction is to investigate parity-odd
contributions to the stationary thermal effective
action, including anomaly-induced terms, and
understand how they modify both the high-temperature
rotating-sphere data and the corresponding
twisted R\'enyi coefficients.
Finally, locality imposes nontrivial relations
among high-temperature CFT coefficients
~\cite{FromZtoa}. By Eq.~\eqref{eq:physicalmapmt},
any such relations involving $a_{0,q;0}^{\rm trace}$
already imply corresponding relations among the local
R\'enyi coefficients. A systematic classification
of these constraints remains of interest.

We have focused on the local
thermal-EFT contributions to twisted R\'enyi entropies.
A systematic treatment of additional contributions,
including rotational winding sectors~\cite{AngularFractals}
and gauge zero and edge modes~\cite{DonnellyWall,SoniTrivedi2016},
is an interesting direction for future work.

\subsection{Acknowledgements} The author thanks Christopher Herzog for comments on the manuscript.
Claude Opus $5.5$ (Anthropic) was used to assist with
scientific discussions, manuscript drafting, literature searches,
and verification of formulae.
The author takes full responsibility for the scientific
content and accuracy of the manuscript.


\nocite{DeserSchwimmer,GiombiMaloneyYin,MukherjeeWorkInProgress}

\putbib[references]


\appendix

\section*{APPENDIX A: Weyl-invariant reduction on the hyperbolic cylinder}
\refstepcounter{prlappendix}\label{appA}

In this appendix, we explain why the invariant densities entering the
local correspondence have the same functional form
on the two geometries. The thermal identification on
the hyperbolic cylinder is generated by
$v=2\pi n\,\partial_\tau+\theta\,\partial_\phi$.
Its norm, which determines the local proper length of
the thermal circle, is
\begin{equation}
 v^2=(2\pi n)^2+\theta^2\sinh^2r.
 \label{eq:appendixnorm}
\end{equation}

The all-order correspondence has a simple geometric
origin. Let us consider the metric on the rotating-sphere
cylinder, with thermal identification vector evaluated
on the formal $\beta_S=0$ slice,
\begin{equation}
\begin{aligned}
ds_S^2&=d\eta^2+d\psi^2
+\sin^2\psi\,d\tau^2+\cos^2\psi\,d\phi^2,\\
v_S&=(2\pi n)\partial_\tau+\theta\partial_\phi.
\end{aligned}
\end{equation}
Under the coordinate transformation
\begin{equation}
\sin\psi=\frac{1}{\cosh r},
\qquad
\cos\psi=\tanh r,
\end{equation}
the rotating-sphere and hyperbolic-cylinder metrics
are locally related by
\begin{equation}
g_S=\frac{1}{\cosh^2r}\,g_H,
\qquad v_S=v_H.
\label{eq:appendixWeyl}
\end{equation}
Thus, the two metric--thermal-vector pairs are
locally Weyl equivalent. Consequently, every
Weyl-invariant local thermal density agrees on
the two geometries, independently of its derivative
order. In particular, the measure
$\sqrt g/(v^2)^2$ is Weyl invariant in four dimensions.
This establishes the all-order matching of the local
Weyl-invariant bulk sector without requiring an
explicit classification of independent scalar invariants.

The conformal equivalence is local. The relative
normalization of the integrated actions is established
in Appendix~\ref{appB}, while anomalous Weyl
contributions are treated separately in
Appendix~\ref{appC}.

To describe the thermal circle as a fibration, we choose a
unit-period coordinate $t=\tau/(2\pi n)$ and
$\varphi=\phi-\theta t$. The Killing vector becomes
$\partial_t$, and the metric can be written as
\begin{align}
 ds^2&=v^2(dt+\mathcal A)^2+h_{ij}dx^idx^j,\nonumber\\
 \mathcal A&=\frac{\theta\sinh^2r}{v^2}\,d\varphi,
 \label{eq:appendixKK}\\[-6pt]
 h&=dr^2+\cosh^2r\,d\eta^2
  +\frac{(2\pi n)^2\sinh^2r}{v^2}d\varphi^2.
 \nonumber
\end{align}
Under a Weyl transformation, $h$ and $v^2$ have
the same weight. Hence the quotient metric $h/v^2$
and the one-form $\mathcal A$ encode Weyl-invariant
local data. In particular, their curvature and
field strength, written using the variable $z$ in
Eq.~\eqref{eq:zvar}, obey
\begin{equation}
 \mathcal R[h/v^2]=20z-6[(2\pi n)^2+\theta^2],
 \qquad (d\mathcal A)^2=8z.
 \label{eq:appendixcurv}
\end{equation}
Here the last contraction uses the metric $h/v^2$.
The Weyl-covariant combinations used in the main
text consequently take the polynomial form displayed
in Eq.~\eqref{eq:hyperinv}. For instance, the two
independent scalars at two-derivative order reduce
to $-z$ and $3z-(2\pi n)^2-\theta^2$.

These expressions illustrate the reduction of local
thermal invariants to functions of the radial variable
$z$ and the angular displacements. More generally,
the matching of higher-derivative Weyl-invariant
densities follows directly from the local conformal
equivalence in Eq.~\eqref{eq:appendixWeyl}, without
assuming that the displayed invariants exhaust the
independent scalar structures at every derivative
order.


\section*{APPENDIX B: Reduced measures and the coefficient relation}
\refstepcounter{prlappendix}\label{appB}

We now verify the normalization in
Eq.~\eqref{eq:Gidentity} directly. For a fixed local
thermal density, let $\mathcal I(z)$ denote its value
on the hyperbolic cylinder after inserting the
invariants in Eq.~\eqref{eq:hyperinv}. Its radial
integral is Eq.~\eqref{eq:hyperradial}. Differentiating
Eq.~\eqref{eq:zvar}, we obtain
\begin{equation}
 \frac{\sinh r\cosh r\,dr}{(v^2)^2}
 =\frac{dz}{2(2\pi n)^2\theta^2
 [(2\pi n)^2-\theta^2]}.
 \label{eq:appendixjac}
\end{equation}
At $r=0$ and $r=\infty$, respectively,
$z=\theta^2$ and $z=(2\pi n)^2$. Performing the
integrals over the twisted circle and $\phi$ gives
\begin{equation}
 \begin{aligned}
 \log Z_{\HH^3}^{\loc}
 &=\frac{\pi(2\pi n)L}
 {(2\pi n)^2\theta^2[(2\pi n)^2-\theta^2]}
 \times\int_{\theta^2}^{(2\pi n)^2}
 dz\,\mathcal I(z).
 \end{aligned}
 \label{eq:appendixH}
\end{equation}

For the rotating sphere, write the unit-round
metric as
$ds_{S^3}^2=d\psi^2+\sin^2\psi\,d\phi_1^2
+\cos^2\psi\,d\phi_2^2$.
On the formal $\beta=0$ slice, the thermal vector
has angular displacements $2\pi n$ and $\theta$.
Its squared norm is
$w_S=(2\pi n)^2\sin^2\psi
+\theta^2\cos^2\psi$.
The variable
$z_S=(2\pi n)^2\theta^2/w_S$ gives exactly the
same invariant density $\mathcal I(z_S)$ by
conformal covariance. Integrating over both
sphere angles and substituting $z_S$ yields
\begin{equation}
 \begin{aligned}
&\left.\frac{\log Z_{S^3}^{\rot}}{\beta}
 \right|_{\substack{\beta=0\\
 \alpha_1=2\pi n,\,\alpha_2=\theta}}\\
&\qquad=\frac{2\pi^2}
 {(2\pi n)^2\theta^2[(2\pi n)^2-\theta^2]}
 \int_{\theta^2}^{(2\pi n)^2}dz\,\mathcal I(z).
\end{aligned}
 \label{eq:appendixS}
\end{equation}
The ratio of the prefactors in
Eqs.~\eqref{eq:appendixH} and \eqref{eq:appendixS}
is $nL$: this proves Eq.~\eqref{eq:Gidentity}
without using any particular thermal-EFT
coefficient. The equality is understood order by
order under the indicated formal continuation.

At derivative order $2j$, homogeneity implies
$\mathcal I_j(z;0,\theta^2)
=\theta^{2j}\mathcal I_j(z/\theta^2;0,1)$.
Taking $n\to0$ in Eq.~\eqref{eq:appendixH},
we obtain
\begin{equation}
 \left.\log Z_{\HH^3}^{\loc,(2j)}\right|_{L/n}
 =\frac{L}{2n}\,\theta^{2j-2}
 \int_0^1dt\,\mathcal I_j(t;0,1).
 \label{eq:appendixprojection}
\end{equation}
Thus, the local twisted R\'enyi coefficients
are determined by the corresponding rotating-sphere
thermal-EFT couplings. Anomalous Weyl contributions
and nonuniform rotational windings are treated separately.


\section*{APPENDIX C: Finite Weyl anomaly in the local thermal frame}
\refstepcounter{prlappendix}\label{appC}

We derive the anomalous correction appearing in
Eq.~\eqref{eq:hyperanomaly} directly on the
hyperbolic cylinder. The Weyl-invariant thermal
effective action is naturally evaluated in the
unit-thermal-length frame,
$\widehat g_H=g_H/v^2$, where
$v^2=\beta_H^2+\theta^2\sinh^2r$ and
$\beta_H=2\pi n$.
The corresponding Weyl factor, up to a constant
rescaling, is
\begin{equation}
 \sigma(r)=\frac12\log\left(
 1+\frac{\theta^2}{\beta_H^2}\sinh^2r
 \right).
 \label{eq:appCpath}
\end{equation}
In the $b=0$ counterterm scheme, both frames are
conformally flat, so only the type-A anomaly
contributes. We use the finite Wess--Zumino
functional~\cite{SchwimmerTheisen}, whose
explicit form and evaluation are given in the
Supplemental Material~\cite{Supplement}.

In the leading small-$n$ limit, the local
contribution is concentrated near the rotational
fixed axis, $r\sim\beta_H/|\theta|$.
Introducing $w=\theta^2r^2/\beta_H^2$,
the transverse integral of the finite
Wess--Zumino action reduces to
\begin{equation}
 \begin{aligned}
 \Delta_H
 &=-\frac{a_{\rm W}}{16\pi^2}
 \frac{2\pi L\theta^2}{\beta_H}
 \int_0^\infty dw\,
 \frac{w(w-4)}{(1+w)^4}
 +o(L/\beta_H)\\
 &=\frac{L}{n}
 \frac{a_{\rm W}\theta^2}{48\pi^2}
 +o(L/n).
 \end{aligned}
 \label{eq:appCresult}
\end{equation}
This reproduces the leading small-$n$ projection
of the rotating-sphere anomaly in
Eq.~\eqref{eq:sphereanomaly}, establishing the
correspondence for the physical thermal
coefficients in Eq.~\eqref{eq:physicalmapmt}.

The holographic calculation involves a different
Weyl transformation, relating the natural
black-hole boundary metric
$g_b=D^{-1}g_H$, with
$D=1+\alpha^2\sinh^2r$, to $g_H$.
In the leading small-$n$ limit, the horizon
relations give
\begin{equation}
 \theta=-\frac{2\pi q}{2-q^2},
 \qquad
 \alpha=-\frac{\theta}{\beta_H}(1-q^2),
 \qquad q=\frac{\alpha}{r_+}.
 \label{eq:appCgravityrelation}
\end{equation}
The same finite Wess--Zumino functional,
now evaluated with the Weyl factor determined
by $D$, gives
\begin{equation}
 \begin{aligned}
 \left.\Delta_{b\to H}\log Z\right|_{L/n}
 &=\frac{L}{n}\frac{a_{\rm W}}{48\pi^2}
 \theta^2(1-q^2)^2\\
 &=\frac{L}{n}N^2\left[
 \frac{\theta^2}{192\pi^2}
 -\frac{\theta^4}{96\pi^4}
 +\ord(\theta^6)\right],
 \end{aligned}
 \label{eq:appCgravWZ}
\end{equation}
where $a_{\rm W}=N^2/4$.
Thus, the additional $\theta^4$ term in the
holographic calculation arises because the
two Wess--Zumino contributions connect
different conformal frames.

\end{bibunit}


\clearpage
\onecolumngrid

\setcounter{equation}{0}
\renewcommand{\theequation}{S\arabic{equation}}

\begin{bibunit}[apsrev4-2]

\section*{Supplemental Material}
\label{sec:supplemental}

In this supplemental material, we provide the details of the independent holographic and free-field checks of the local
perturbative coefficient relation described in the Letter.

The angular-momentum twisted R\'enyi partition function of a conformal field theory across a spherical entangling surface can be computed using Casini-Huerta-Myers map~\cite{CHM}
\begin{align}
 Z_n(\theta)&=\Tr_{\HH^3}\!\left(e^{-\beta H}e^{i\theta J}\right),
 &\beta&=2\pi n,\nonumber\\
 ds_H^2&=d\tau^2+dr^2+\cosh^2r\,d\eta^2+\sinh^2r\,d\phi^2,
 & (\tau,\phi)&\sim(\tau+\beta,\phi+\theta).
 \label{eq:shmetric}
\end{align}
Introducing a UV cutoff \(\epsilon\) near the spherical entangling surface in the original flat-space geometry induces the following cutoff on hyperbolic space:
\begin{equation}
 \cosh r\cosh\eta\leq\cosh U,\qquad
 U=\log\frac{2R}{\epsilon}+\ord(\epsilon/R),\qquad
 L=2U=2\log\frac R\epsilon+\ord(1).
 \label{eq:scut}
\end{equation}
We develop a thermal effective action on the hyperbolic cylinder and the local expansion of the effective field theory is organized for $\beta=2\pi n\ll|\theta|\ll1$.
We define $K^{\pert}(\theta)$ by extracting the coefficient of $L/n$
from the local thermal effective action, or, in the free-field winding
representation, by the corresponding termwise Mellin/$\zeta$-function
coefficient prescription:
\begin{align}
 \left.\log Z_n(\theta)\right|_{\loc,\log,\pert}
 &=\frac L n K^{\pert}(\theta)+\text{subleading local terms in }n,\nonumber\\
 \left.n S_n(\theta)\right|_{\loc,\log,\pert}
 &=L K^{\pert}(\theta)+\text{subleading local terms in }n.
 \label{eq:sKdef}
\end{align}
Here $S_n(\theta)=(1-n)^{-1}\log[\mathcal Z_n(\theta)/\mathcal Z_1(\theta)^n]$, with $\mathcal Z_n(\theta)=Z_n(\theta)/Z_1(0)^n$. The normalization terms do not affect the leading $L/n$ coefficient.
We work throughout in the local thermal-EFT regime
$2\pi n\ll|\theta|\ll1$, where the perturbative expansion is organised by first taking the small-$n$ limit of the local contribution and subsequently expanding in small $\theta$.

\suppsubheading{All-order correspondence with the rotating-sphere EFT}

We write the metric on $S^1\times\HH^3$ in cylindrical
coordinates as
\begin{equation}
 ds^2=d\tau^2+dr^2+\cosh^2r\,d\eta^2
       +\sinh^2r\,d\phi^2, \qquad
(\tau,\eta,\phi)
\sim(\tau+\beta,\eta,\phi+\theta).
\end{equation}
The thermal vector and its norm are
\begin{equation}
 v=\beta\partial_\tau+\theta\partial_\phi,
 \qquad v^2=\beta^2+\theta^2\sinh^2r.
 \label{eq:sThermalNorm}
\end{equation}
We introduce the variable
\begin{equation}
 z=\frac{\beta^2\theta^2\cosh^2r}
 {\beta^2+\theta^2\sinh^2r}.
 \label{eq:sZvariable}
\end{equation}
Using the unit-thermal-norm Weyl frame
$\widehat g=g/v^2$, the vorticity and
Weyl-covariant Schouten tensors give
\begin{align}
 \frac18\Tr\omega^2&=-z,\nonumber\\
 v^2\mathcal S&=3z-\beta^2-\theta^2,\nonumber\\
 v^2\mathcal B_\mu(\omega^2)^{\mu\nu}\mathcal B_\nu
 &=4z(z-\beta^2)(z-\theta^2),
 \qquad
 \mathcal B_\mu=\mathcal S_{\mu\nu}v^\nu.
 \label{eq:sHdata}
\end{align}
The first relation follows from
$\omega=d(v^\flat/v^2)$ in the unit-norm frame,
the second from its scalar curvature, and the third
from the conformal-Killing identity
$\omega_\mu{}^\nu\mathcal B_\nu=-v^2D_\mu\mathcal S/3$.

We now consider the local thermal-EFT integral.
The $\eta$ integral along the CHM-regulated fixed
geodesic gives the logarithmic factor $L$ in
Eq.~\eqref{eq:scut}. The remaining transverse integral
can be written in terms of $z$ using
\begin{equation}
 \frac{\sinh r\cosh r\,dr}{(v^2)^2}
 =\frac{dz}{2\beta^2\theta^2(\beta^2-\theta^2)}.
 \label{eq:sMeasure}
\end{equation}
The integration limits are $z=\theta^2$ and
$z=\beta^2$, with the orientation understood when
$\beta^2<\theta^2$.

We next compare this integral with the rotating-sphere
EFT on $S^1\times S^3$. We write the round
three-sphere metric as
\begin{equation}
 ds_{S^3}^2=d\psi^2+\cos^2\psi\,d\phi_1^2
                 +\sin^2\psi\,d\phi_2^2,
 \qquad 0\leq\psi\leq\frac\pi2.
\end{equation}
Taking the two rotational displacements to be
$(\beta,\theta)$, the rotational vector and its norm,
 are
\begin{align}
 v_S&=\beta\partial_{\phi_1}
       +\theta\partial_{\phi_2},\nonumber\\
 v_S^2&=\beta^2\cos^2\psi
       +\theta^2\sin^2\psi.
\end{align}
Introducing, the variable $z_S$
\begin{equation}
 z_S=\frac{\beta^2\theta^2}{v_S^2},
\end{equation}
we find that the three scalar quantities in
Eq.~\eqref{eq:sHdata} take exactly the same form
on the rotating sphere, with $z$ replaced by $z_S$.
In particular, if the sphere basis is written using
$\zeta=\Tr\omega^2/8$, our convention gives
$\zeta=-z_S$, fixing the relative sign.

The sphere integral also takes the same form,
since
\begin{equation}
 \frac{\sin\psi\cos\psi\,d\psi}{(v_S^2)^2}
 =\frac{dz_S}
 {2\beta^2\theta^2(\beta^2-\theta^2)}.
\end{equation}
The integration limits are again $\theta^2$ and
$\beta^2$. In fact, the change of coordinates
\begin{equation}
 \tan\psi=\sinh r
\end{equation}
identifies $z_S=z$ and the two reduced measures.

Since every relevant parity-even Weyl-invariant bulk
density can be expressed as a polynomial in these
three scalars, up to total derivatives
\cite{FromZtoa}, the two backgrounds
give the same reduced local integrals.
The transverse angular integrals give $2\pi$
on the hyperbolic cylinder and $(2\pi)^2$
on the sphere. We therefore obtain the
all-order relation
\begin{equation}
 \log Z_{H}^{\rm WI,loc}(n,\theta)
 =nL\left.
 \frac{\log Z_{S^3}^{\rm WI,rot}
 (\beta_S,\alpha_1,\alpha_2)}{\beta_S}
 \right|_{\substack{
 \beta_S=0\\
 \alpha_1=2\pi n,\;\alpha_2=\theta}}.
 \label{eq:sGidentity}
\end{equation}
Here the formal $\beta_S=0$ specialization and
the $n\to0$ limit are understood as coefficient
extractions in the local derivative expansion.
Neither defines a pointwise angular thermal trace.

\suppsubheading{Wess--Zumino contribution for a general
four-dimensional CFT}

The Weyl-invariant thermal effective action is naturally
evaluated in the frame where the thermal vector has
unit norm. On the hyperbolic cylinder, this frame is
defined by
\begin{equation}
 \widehat g_H=\frac{g_H}{v^2},
 \label{eq:sUnitFrame}
\end{equation}
where the thermal vector and its norm are given in
Eq.~\eqref{eq:sThermalNorm}.

In even dimensions, the partition function is not
invariant under a Weyl transformation because of
the Weyl anomaly. Under
$g_{ij}\to e^{2\omega}g_{ij}$, its variation in $d=4$ dimension is
given by~\cite{DeserSchwimmer,SchwimmerTheisen}
\begin{equation}
 \delta_\omega\log Z[g]
 =\frac{1}{16\pi^2}
 \int d^4x\sqrt g\,\omega
 \left(
 cW_{ijkl}W^{ijkl}
 -a_{\rm W}E_4
 +b\Box R
 \right),
 \label{eq:sAnomalyGen}
\end{equation}
where $W_{ijkl}$ is the Weyl tensor and
$E_4$ is the four-dimensional Euler density,
\begin{equation}
 E_4=R_{ijkl}R^{ijkl}-4R_{ij}R^{ij}+R^2.
\end{equation}
The coefficient $b$ is scheme dependent and can
be shifted by a finite local $R^2$ counterterm.
We work in the scheme $b=0$
\cite{DeserSchwimmer,Skenderis}.

Since the hyperbolic cylinder is conformally flat,
the Weyl tensor vanishes in both frames.
Therefore, only the Euler anomaly contributes.
We write the Weyl transformation as
\begin{equation}
 \widehat g_H=\beta^{-2}e^{-2\sigma}g_H,
 \qquad
 \sigma(r)=\frac12\log\left(
 1+\frac{\theta^2}{\beta^2}\sinh^2r
 \right).
 \label{eq:sWeylFactor}
\end{equation}
The constant rescaling by $\beta^{-2}$ does not
contribute to the leading local $L/\beta$ term
in the chosen scheme.

The finite Wess--Zumino contribution associated
with this transformation is
\cite{SchwimmerTheisen}
\begin{align}
 \Delta_{\rm th\to H}\log Z
 =-\frac{a_{\rm W}}{16\pi^2}
 \int d^4x\sqrt{g_H}\,
 \big[
 &\sigma E_4[g_H]
 +4G_H^{ij}\partial_i\sigma\partial_j\sigma
 -4(\nabla\sigma)^2\nabla^2\sigma
 +2(\nabla\sigma)^4
 \big],
 \label{eq:sWZAction}
\end{align}
where $G_H^{ij}=R_H^{ij}-g_H^{ij}R_H/2$
is the Einstein tensor of $g_H$.

For $S^1\times\HH^3$, we have
\begin{equation}
 E_4[g_H]=0,
 \qquad
 G_H^{rr}=1.
\end{equation}
Since $\sigma$ depends only on $r$,
\begin{equation}
 (\nabla\sigma)^2=(\sigma')^2,
 \qquad
 \nabla^2\sigma
 =\sigma''+(\tanh r+\coth r)\sigma'.
\end{equation}

We now evaluate the leading contribution in the
limit $\beta=2\pi n\to0$ at fixed nonzero $\theta$.
The dominant local contribution comes from
$r\sim\beta/|\theta|$, near the rotational fixed axis.
In this region, the metric becomes
\begin{equation}
 ds_H^2\simeq
 d\tau^2+d\eta^2+dr^2+r^2d\phi^2.
\end{equation}
Introducing the variable
\begin{equation}
 w=\frac{\theta^2r^2}{\beta^2},
 \qquad
 \sigma=\frac12\log(1+w),
 \label{eq:sScaledVariable}
\end{equation}
we obtain
\begin{equation}
 (\nabla\sigma)^2
 =\frac{\theta^2}{\beta^2}
 \frac{w}{(1+w)^2},
 \qquad
 \nabla^2\sigma
 =\frac{2\theta^2}{\beta^2}
 \frac{1}{(1+w)^2}.
\end{equation}

The term involving the Einstein tensor is
subleading in the small-$\beta$ expansion.
The leading contribution therefore comes from
the last two terms in Eq.~\eqref{eq:sWZAction},
which give
\begin{equation}
 -4(\nabla\sigma)^2\nabla^2\sigma
 +2(\nabla\sigma)^4
 =
 \frac{2\theta^4}{\beta^4}
 \frac{w(w-4)}{(1+w)^4}.
\end{equation}

Using the transverse measure
$d^2x_\perp=r\,dr\,d\phi$, with
$
 r\,dr=\frac{\beta^2}{2\theta^2}\,dw,
$
we find
\begin{align}
 &\int d^2x_\perp
 \left[
 -4(\nabla\sigma)^2\nabla^2\sigma
 +2(\nabla\sigma)^4
 \right]
 =
 \frac{2\pi\theta^2}{\beta^2}
 \int_0^\infty dw\,
 \frac{w(w-4)}{(1+w)^4}
 =
 -\frac{2\pi\theta^2}{3\beta^2}.
 \label{eq:sUnivInt}
\end{align}

Including the thermal period $\beta$ and the
regulated length $L$ along the fixed geodesic,
Eq.~\eqref{eq:sWZAction} gives
\begin{align}
 \Delta_{\rm th\to H}\log Z
 &=
 \frac{a_{\rm W}L\theta^2}{24\pi\beta}
 +o(L/\beta)=
 \frac{L}{n}
 \frac{a_{\rm W}\theta^2}{48\pi^2}
 +o(L/n).
 \label{eq:sUnivShift}
\end{align}

Thus, the leading local Weyl-anomaly contribution
is determined without expanding in $\theta$.
In particular, it contributes only at order
$\theta^2$,
\begin{equation}
 \Delta K_{\rm th\to H}(\theta)
 =
 \frac{a_{\rm W}\theta^2}{48\pi^2}.
 \label{eq:sAnomalyK}
\end{equation}
There are no additional analytic
$\theta^4,\theta^6,\ldots$ contributions to this
leading local term.

This result concerns the Weyl transformation
between the unit-thermal-norm frame and the
physical hyperbolic cylinder. It is distinct
from the finite Weyl transformation between
the natural conformal boundary of the rotating
AdS$_5$ solution and the hyperbolic cylinder,
which is evaluated separately in the
holographic calculation.

\suppsubheading{Agreement with the spherical anomaly and coefficient map}

We now compare the Wess--Zumino contribution obtained
above with the anomaly contribution in the rotating-sphere
EFT. In the conventions of Ref.~\cite{FromZtoa},
the finite type-A anomaly shift between the trace and
Weyl-invariant frames is
\begin{equation}
 \frac{\log Z_{S^3}^{\rm trace}
 -\log Z_{S^3}^{\rm WI,rot}}{\beta_S}
 =\frac{a_{\rm W}}{12}
 \left(\frac{y_1}{y_2}+\frac{y_2}{y_1}+7\right),
 \qquad
 y_i=\beta_S^2+\alpha_i^2.
 \label{eq:sSphereAnomaly}
\end{equation}
Setting $\beta_S=0$, $\alpha_1=2\pi n$ and
$\alpha_2=\theta$, and multiplying by $nL$, we obtain
\begin{align}
 nL\left.
 \frac{\log Z_{S^3}^{\rm trace}
 -\log Z_{S^3}^{\rm WI,rot}}{\beta_S}
 \right|_{\substack{\beta_S=0\\
 \alpha_1=2\pi n,\,\alpha_2=\theta}}=
 \frac{a_{\rm W}nL}{12}
 \left(
 \frac{4\pi^2n^2}{\theta^2}
 +\frac{\theta^2}{4\pi^2n^2}+7
 \right)
 =
 \frac{L}{n}\frac{a_{\rm W}\theta^2}{48\pi^2}
 +\ord(nL).
 \label{eq:sSphereAnomMatch}
\end{align}
This agrees precisely with the Wess--Zumino contribution
obtained on the hyperbolic cylinder in
Eq.~\eqref{eq:sUnivShift}.

Denoting the operator-counting coefficients in the
Weyl-invariant and trace frames by
$a_{0,k;0}^{\rm WI}$ and $a_{0,k;0}^{\rm trace}$,
respectively, the anomaly shifts the coefficients
according to
\begin{equation}
 a_{0,k;0}^{\rm trace}-a_{0,k;0}^{\rm WI}
 =-\frac{a_{\rm W}}{12}\delta_{k,2},
 \qquad
 s_k^{\rm phys,loc}-s_k^{\rm WI}
 =\frac{a_{\rm W}}{48\pi^2}\delta_{k,2}.
 \label{eq:sCoefficientShift}
\end{equation}
Combining these relations with the Weyl-invariant
correspondence in Eq.~\eqref{eq:sGidentity},
we obtain
\begin{equation}
 s_k^{\rm phys,loc}
 =-\frac{a_{0,k;0}^{\rm trace}}{4\pi^2},
 \qquad k\ge1.
 \label{eq:sPhysicalMap}
\end{equation}
This establishes the all-order correspondence between
the local logarithmic coefficients of the twisted
R\'enyi entropy and the rotating-sphere EFT coefficients,
including the Weyl-anomaly contribution.


\suppheading{Independent holographic test at large $N$}

\emph{Exact continued topological Kerr--AdS$_5$ geometry.}---
We work in five-dimensional Einstein gravity, with
the AdS radius set to one,
\begin{equation}
 R_{MN}+4G_{MN}=0,\qquad
 G_5^{-1}=\frac{2N^2}{\pi},\qquad
 a_{\rm W}=c=\frac{\pi}{8G_5}=\frac{N^2}{4}.
 \label{eq:sgravitynorm}
\end{equation}
The bulk dual of the thermal CFT on $\HH^3$ is the
topological AdS$_5$ black hole. We consider its
singly rotating generalization
\cite{Dehghani,HeRotating}, analytically continued
from the Lorentzian rotation parameter $a_L=i\alpha$.
The Euclidean metric takes the form
\begin{align}
 ds_5^2={}&\frac{\Delta}{P D^2}
 \bigl(d\tau-\alpha\sinh^2r\,d\phi\bigr)^2
 +\frac P\Delta\,d\hat r^2+\frac P D\,dr^2
 \nonumber\\
 &+\frac{\sinh^2r}{P D^2}
 \bigl[\alpha(1-\hat r^2)d\tau
 -(\hat r^2-\alpha^2)d\phi\bigr]^2
 +\frac{\hat r^2\cosh^2r}{D}\,d\eta^2,
 \label{eq:skerrmetric}
\end{align}
where
\begin{equation}
 D=1+\alpha^2\sinh^2r,\qquad
 P=\hat r^2-\frac{\alpha^2\cosh^2r}{D},\qquad
 \Delta=(\hat r^2-\alpha^2)(\hat r^2-1)-2m,
 \label{eq:smetricfunctions}
\end{equation}
Here $\hat r$ is the radial coordinate in the
five-dimensional bulk, while $r$ parametrizes
the transverse direction of the boundary hyperbolic
space. The horizon is located at $\hat r=r_+$,
where $r_+$ is the largest root of $\Delta=0$,
with $r_+>\max(1,|\alpha|)$. Thus, the location of the horizon can be found from the following equation
\begin{equation}
 2m=(r_+^2-\alpha^2)(r_+^2-1).
\end{equation}
The determinant of the bulk metric is also given by
\begin{equation}
 \sqrt G=\frac{\hat r P\sinh r\cosh r}{D^2}.
 \label{eq:sBulkDet}
\end{equation}
The natural conformal boundary metric is obtained
by taking $\hat r\to\infty$,
\begin{equation}
 g_b=\lim_{\hat r\to\infty}
 \hat r^{-2}G\big|_{d\hat r=0}
 =\frac{g_H}{D},
 \qquad D=1+\alpha^2\sinh^2r.
 \label{eq:snaturalframe}
\end{equation}
The finite Weyl-anomaly contribution associated
with the transformation from $g_b$ to the physical
hyperbolic metric $g_H$ must therefore be included
in the renormalized on-shell action.

Regularity at the horizon determines the thermal
period $\beta$ and angular twist $\theta$.
The horizon angular velocity and the corresponding
identifications are
\begin{align}
 \Omega_H&=-\frac{\alpha(r_+^2-1)}{r_+^2-\alpha^2},
 \nonumber\\
 \beta&=\frac{2\pi(r_+^2-\alpha^2)}
 {r_+(2r_+^2-\alpha^2-1)},
 \nonumber\\
 \theta&=\beta\Omega_H
 =-\frac{2\pi\alpha(r_+^2-1)}
 {r_+(2r_+^2-\alpha^2-1)}.
 \label{eq:shorizon}
\end{align}

For the high-temperature expansion, we introduce
\begin{equation}
 q=\frac{\alpha}{r_+},
 \qquad
 \kappa=r_+^{-2}.
\end{equation}
The relations in Eq.~\eqref{eq:shorizon} become
\begin{equation}
 \beta r_+
 =\frac{2\pi(1-q^2)}{2-q^2-\kappa},
 \qquad
 \theta
 =-\frac{2\pi q(1-\kappa)}{2-q^2-\kappa}.
 \label{eq:sqrelation}
\end{equation}
We take the high-temperature limit
$r_+\to\infty$ with $q$ fixed, so that
$\alpha=qr_+$ and the physical twist $\theta$
remains finite. In this limit,
\begin{equation}
 \beta r_+
 =\frac{2\pi(1-q^2)}{2-q^2}
 +\ord(\beta^2),
 \qquad
 q=-\frac{\theta}{\pi}
 +\frac{\theta^3}{2\pi^3}
 +\ord(\theta^5).
 \label{eq:sqseries}
\end{equation}
We first extract the leading high-temperature
contribution at fixed $\theta$ and then expand
the result at small $\theta$.

The relevant contribution comes from the region
$r\sim \alpha^{-1}\sim\beta/|\theta|$ near the
rotational axis. We use the exact continued
metric to determine the local high-temperature
coefficient, without assuming a smooth global
completion of the twisted boundary geometry.
The corresponding cutoff estimate is given
after Eq.~\eqref{eq:sInheritedLength}.

\suppsubheading{Einstein, boundary, and counterterm actions}

We evaluate the renormalized Euclidean action using
standard holographic renormalization
\cite{BalasubramanianKraus,Skenderis,HenningsonSkenderis}.
The action consists of the Einstein--Hilbert term,
the Gibbons--Hawking boundary term, and the
counterterms,
\begin{equation}
 I_E=I_{\EH}+I_{\GH}+I_{\ct},
\end{equation}
where
\begin{align}
 I_{\EH}
 &=-\frac{1}{16\pi G_5}
 \int_M d^5x\,\sqrt G\,(R+12),
 \nonumber\\
 I_{\GH}
 &=-\frac{1}{8\pi G_5}
 \int_{\partial M}d^4x\,\sqrt h\,K,
 \nonumber\\
 I_{\ct}
 &=\frac{1}{8\pi G_5}
 \int_{\partial M}d^4x\,\sqrt h
 \left(3+\frac14\mathcal R[h]\right)
 +\frac{\log(\Lambda/\mu)}{64\pi G_5}
 \int_{\partial M}d^4x\,\sqrt h
 \left(
 \mathcal R_{ab}[h]\mathcal R^{ab}[h]
 -\frac13\mathcal R[h]^2
 \right).
 \label{eq:scaction}
\end{align}
Here $h_{ab}$ is the induced boundary metric,
$K$ is the trace of the extrinsic curvature,
and $\mathcal R_{ab}[h]$ is the Ricci tensor
of $h_{ab}$.

We introduce a large radial cutoff
$\hat r=\Lambda$. On this surface, the induced
metric is
\begin{align}
 ds_h^2={}&
 \frac{\Delta_\Lambda}{P_\Lambda D^2}
 \bigl(d\tau-\alpha\sinh^2r\,d\phi\bigr)^2
 +\frac{P_\Lambda}{D}\,dr^2
 +\frac{\sinh^2r}{P_\Lambda D^2}
 \bigl[\alpha(1-\Lambda^2)d\tau
 -(\Lambda^2-\alpha^2)d\phi\bigr]^2
+\frac{\Lambda^2\cosh^2r}{D}\,d\eta^2,
 \label{eq:sInducedMetric}
\end{align}
where $P_\Lambda=P|_{\hat r=\Lambda}$
and $\Delta_\Lambda=\Delta|_{\hat r=\Lambda}$.
We introduce a regulated length $L_\eta$
along the fixed geodesic to extract the
local tube contribution. The relation to
the CHM cutoff is discussed below.

Using the on-shell relation $R=-20$ and the
bulk determinant in Eq.~\eqref{eq:sBulkDet},
the Einstein--Hilbert contribution becomes
\begin{equation}
 I_{\EH}
 =\frac{\beta L_\eta}{8G_5\alpha^2}
 \left[
 \Lambda^4-r_+^4
 -(\alpha^2+1)(\Lambda^2-r_+^2)
 \right].
 \label{eq:sehresult}
\end{equation}
The Gibbons--Hawking term evaluated on the
cutoff surface gives
\begin{align}
 I_{\GH}
 =-\frac{\beta L_\eta}{4G_5}
 \bigg[
 &\frac{2\Lambda^4}{\alpha^2}
 -\frac{5(\alpha^2+1)\Lambda^2}{4\alpha^2}
 -\frac{
 \alpha^4-12\alpha^2r_+^2+4\alpha^2
 +12r_+^4-12r_+^2+1
 }{12\alpha^2}
 \bigg]
 +\ord(\Lambda^{-2}).
 \label{eq:sGHresult}
\end{align}
Similarly, the power counterterms give
\begin{align}
 I_{\ct}^{\rm power}
 =\frac{\beta L_\eta}{G_5}
 \bigg[
 &\frac{3\Lambda^4}{8\alpha^2}
 -\frac{3(\alpha^2+1)\Lambda^2}{16\alpha^2}
-\frac{\alpha^4-5\alpha^2+24m+1}
 {64\alpha^2}
 \bigg]
 +\ord(\Lambda^{-2}).
 \label{eq:sctresult}
\end{align}
The $\Lambda^4$ and $\Lambda^2$ divergences
cancel upon adding these three contributions.
The logarithmic counterterm has vanishing
integrated contribution to the leading local
 term.

Using the horizon relation
$2m=(r_+^2-\alpha^2)(r_+^2-1)$,
we obtain the finite result
\begin{equation}
 I_{\rm ren,tube}^{\rm power}
 =\frac{\beta L_\eta}{192G_5\alpha^2}
 \left[
 \alpha^4+12\alpha^2r_+^2-5\alpha^2
 -12r_+^4+12r_+^2+1
 \right].
 \label{eq:sIren}
\end{equation}
This result includes the Einstein--Hilbert,
Gibbons--Hawking, and power counterterm
contributions. We will evaluate the finite Weyl-anomaly
contribution associated with the transformation
from $g_b$ to $g_H$ separately.

\suppsubheading{Natural-frame high-temperature coefficient}

We now extract the high-temperature coefficients from
the renormalized gravitational action given in \eqref{eq:sIren} and compare them
with the rotating-sphere EFT. The gravitational action
computed above is in the natural boundary frame $g_b$,
whereas the twisted R\'enyi partition function is
defined in the physical hyperbolic frame $g_H$.
We first obtain the coefficients in $g_b$ and then
include the finite Weyl-anomaly contribution below.

Using Eqs.~\eqref{eq:shorizon}--\eqref{eq:sqrelation}
in Eq.~\eqref{eq:sIren}, and taking the
high-temperature limit $\beta\to0$ at fixed $\theta$,
we obtain
\begin{equation}
 \frac{\log Z_b}{L_\eta}
 =\frac{\pi^2}{48G_5\beta}
 \frac{(1-q^2)^2(12-12q^2-q^4)}
 {q^2(2-q^2)^2}
 +\ord(\beta).
 \label{eq:sZbq}
\end{equation}
Expanding this result at small $\theta$ using
Eq.~\eqref{eq:sqrelation}, and setting
$G_5^{-1}=2N^2/\pi$ and $\beta=2\pi n$, we find
\begin{align}
 \left.\frac{n\log Z_b}{L_\eta}\right|_{\beta\to0}
 =N^2\bigg[
 &\frac{\pi^2}{16\theta^2}-\frac1{16}
 +\frac{5\theta^2}{192\pi^2}
 -\frac{\theta^4}{192\pi^4}
 -\frac{\theta^6}{256\pi^6}
 +\ord(\theta^8)
 \bigg].
 \label{eq:sZbcoef}
\end{align}
These are the high-temperature coefficients in the
natural boundary frame. To obtain the corresponding
coefficients in the physical hyperbolic frame, we
must include the finite Weyl-anomaly contribution,
which we compute next.

\suppsubheading{Finite Weyl-anomaly correction}

The gravitational action computed above is defined
in the natural boundary frame $g_b$ as given in \eqref{eq:snaturalframe}. To obtain the
partition function in the physical hyperbolic frame
$g_H$, we must include the finite Weyl-anomaly
contribution. The two metrics are related by
\begin{equation}
 g_b=e^{-2\sigma}g_H,
 \qquad
 \sigma(r)=\frac12\log(1+\alpha^2\sinh^2r).
 \label{eq:sWeylfactor}
\end{equation}
Since both metrics are conformally flat, only the
type-A anomaly contributes in the scheme $b=0$.
The corresponding finite Wess--Zumino action is
given by~\cite{SchwimmerTheisen}
\begin{align}
 \Delta_{b\to H}\log Z
 =-\frac{a_{\rm W}}{16\pi^2}
 \int d^4x\sqrt{g_H}\,
 \big[
 &\sigma E_4[g_H]
 +4G_H^{ij}\partial_i\sigma\partial_j\sigma
 -4(\nabla\sigma)^2\nabla^2\sigma
 +2(\nabla\sigma)^4
 \big],
 \label{eq:sWZidentity}
\end{align}
where $G_H^{ij}$ is the Einstein tensor of $g_H$.

For $S^1\times\HH^3$, we have $E_4[g_H]=0$
and $G_H^{rr}=1$. In the high-temperature limit,
the leading contribution comes from the region
$r\sim\alpha^{-1}$ near the rotational axis.
Introducing
\begin{equation}
 w=\alpha^2r^2,
 \qquad
 \sigma=\frac12\log(1+w),
\end{equation}
we obtain
\begin{equation}
 (\nabla\sigma)^2
 =\frac{\alpha^2w}{(1+w)^2},
 \qquad
 \nabla^2\sigma
 =\frac{2\alpha^2}{(1+w)^2}.
\end{equation}
The Einstein-tensor term is subleading.
The leading transverse integral becomes
\begin{align}
 &\int d^2x_\perp
 \left[
 -4(\nabla\sigma)^2\nabla^2\sigma
 +2(\nabla\sigma)^4
 \right]=
 2\pi\alpha^2
 \int_0^\infty dw\,
 \frac{w(w-4)}{(1+w)^4}
 =-\frac{2\pi\alpha^2}{3}.
 \label{eq:sWZintegral}
\end{align}
Including the thermal period $\beta$ and the
regulated geodesic length $L_\eta$, we find
\begin{equation}
 \Delta_{b\to H}\log Z
 =\frac{a_{\rm W}\beta L_\eta\alpha^2}{24\pi}
 +o(\beta L_\eta\alpha^2).
 \label{eq:sWZfinal}
\end{equation}

Using $\alpha=qr_+$ and the high-temperature
relations in Eq.~\eqref{eq:sqrelation}, this gives
\begin{equation}
 \frac{\Delta_{b\to H}\log Z}{L_\eta}
 =\frac{\pi^2}{48G_5\beta}
 \frac{q^2(1-q^2)^2}{(2-q^2)^2}
 +o(1/\beta).
 \label{eq:sWZcoefficient}
\end{equation}
Expanding in the physical angular twist $\theta$,
we obtain
\begin{equation}
 \left.
 \frac{n\Delta_{b\to H}\log Z}{L_\eta}
 \right|_{\beta\to0}
 =N^2\left[
 \frac{\theta^2}{192\pi^2}
 -\frac{\theta^4}{96\pi^4}
 +\frac{\theta^6}{64\pi^6}
 +\ord(\theta^8)
 \right].
 \label{eq:sWZHigher}
\end{equation}
This correction must be added to the natural-frame
result in Eq.~\eqref{eq:sZbcoef} to obtain the
physical hyperbolic partition function.

Combining the natural-frame result in Eq.~\eqref{eq:sZbq}
with the finite Weyl-anomaly contribution in
Eq.~\eqref{eq:sWZcoefficient}, we obtain
\begin{align}
 \left.\frac{n\log Z_H}{L}\right|_{\beta\to0}
 &=N^2\left[
 \frac{(1-q^2)^2(12-12q^2-q^4)}
 {48q^2(2-q^2)^2}
 +\frac{q^2(1-q^2)^2}{48(2-q^2)^2}
 \right]\nonumber\\
 &=N^2\frac{(1-q^2)^3}
 {4q^2(2-q^2)^2}.
 \label{eq:sfunctionalidentity}
\end{align}
This relation holds for the full leading local
high-temperature function, rather than only
for the first few terms in the small-angle
expansion.

Using the relation between $q$ and the physical
twist $\theta$ in Eq.~\eqref{eq:sqrelation},
and expanding at small $\theta$, we find
\begin{equation}
 K^{\pert}_{\mathcal N=4}(\theta)
 =N^2\left[
 \frac{\pi^2}{16\theta^2}-\frac1{16}
 +\frac{\theta^2}{32\pi^2}
 -\frac{\theta^4}{64\pi^4}
 +\frac{3\theta^6}{256\pi^6}
 +\ord(\theta^8)
 \right].
 \label{eq:slocalresult}
\end{equation}

\suppsubheading{Independent rotating-sphere calculation}

We now compute the rotating-$S^3$ partition function
independently using the thermodynamics of the
singly rotating Kerr--AdS$_5$ black hole.
Here we consider a physical thermal partition function
with one nonzero angular twist.

The Kerr-AdS$_5$ thermodynamic quantities are
given by~\cite{GibbonsPerryPope}
\begin{align}
 \Xi&=1-a_L^2,\qquad
 m=\frac12(r_+^2+a_L^2)(1+r_+^2),
 \nonumber\\
 T&=\frac{r_+(2r_+^2+1+a_L^2)}
 {2\pi(r_+^2+a_L^2)},
 \qquad
 \Omega=\frac{a_L(1+r_+^2)}{r_+^2+a_L^2},
 \nonumber\\
 S&=\frac{\pi^2r_+(r_+^2+a_L^2)}
 {2G_5\Xi},
 \nonumber\\
 M_{\rm state}&=\frac{\pi m(2+\Xi)}
 {4G_5\Xi^2},
 \qquad
 J=\frac{\pi ma_L}{2G_5\Xi^2}.
 \label{eq:sSthermo}
\end{align}
Here $r_+$ denotes the horizon radius of the
spherical black hole. The corresponding
grand-canonical partition function is
\begin{equation}
 \log Z_{S^3}^{\rm state}
 =-\beta(M_{\rm state}-TS-\Omega J)
 =\frac{\pi\beta(r_+^2+a_L^2)(r_+^2-1)}
 {8G_5\Xi}.
 \label{eq:sSlogZ}
\end{equation}

Continuing $a_L=i\alpha$ and taking the
high-temperature limit at fixed physical
twist $\theta$, we obtain
\begin{equation}
 \log Z_{S^3}^{\rm state}
 =\frac{N^2\pi^2}{\beta}
 \frac{(1-q^2)^3}{q^2(2-q^2)^2}
 +\ord(\beta),
 \label{eq:sSphereq}
\end{equation}
where $q=\alpha/r_+$ and its leading relation
to $\theta$ agrees with Eq.~\eqref{eq:sqrelation}.

The rotational fixed circle on the unit sphere
has length $2\pi$. Comparing
Eq.~\eqref{eq:sSphereq} with the hyperbolic
result in Eq.~\eqref{eq:sfunctionalidentity},
we find
\begin{equation}
 \left.\frac{n\log Z_{S^3}^{\rm state}}{2\pi}
 \right|_{\beta\to0}
 =
 \left.\frac{n\log Z_H}{L}\right|_{\beta\to0}
 =
 N^2\frac{(1-q^2)^3}
 {4q^2(2-q^2)^2}.
 \label{eq:sSphereTheta}
\end{equation}
Thus, the two independent gravitational
calculations agree as functions of $q$
at leading order in the high-temperature
expansion, rather than only at a few
orders in $\theta$. Expanding at small
$\theta$ also reproduces
Eq.~\eqref{eq:slocalresult}.

The Casimir energy in the round-sphere
boundary frame is
\begin{equation}
 E_C^{\ESU}=\frac{3\pi}{32G_5}
 =\frac{3N^2}{16}.
 \label{eq:sCasimirRound}
\end{equation}
Its contribution $-\beta E_C^{\ESU}$
is subleading and does not affect the
leading high-temperature coefficients.

\suppsubheading{Cutoff and global regularity}

The cutoff inherited from the CHM map gives
\begin{equation}
 2\eta_{\max}(r)
 =2\operatorname{arcosh}
 \frac{\cosh U}{\cosh r}
 =2U-2\log\cosh r+\cdots.
 \label{eq:sInheritedLength}
\end{equation}
The leading high-temperature contribution comes
from the region $r\sim\alpha^{-1}$ near the
rotational axis. In this region, replacing the
varying geodesic length by $L=2U$ does not change
the leading $L/\beta$ coefficient. The remaining
contributions from the outer region are subleading
in the local radial calculation.

The logarithmic counterterm also has no contribution
to this leading coefficient. In the high-temperature
tube, its transverse integral is proportional to
\begin{equation}
 \int_0^\infty du\,
 \frac{2u-1}{(1+u)^4}=0,
 \qquad u=\alpha^2r^2.
 \label{eq:sLogCTzero}
\end{equation}

Finally, the continued topological Kerr--AdS$_5$
geometry need not be globally regular for the
twisted thermal identification. In this family,
the second shrinking cycle at large $r$ requires
\begin{equation}
 p\beta=2\pi,\qquad
 p\theta+2\pi k=0,
 \qquad p,k\in\mathbb Z.
 \label{eq:sGlobalLattice}
\end{equation}
Smoothness at the intersection of the shrinking
cycles further requires $|k|=1$, implying
$|\theta|=\beta$, which is incompatible with
the regime $\beta\ll|\theta|$.

For our local thermal-EFT calculation, the
near-axis analysis is sufficient to extract the
leading tube contribution, since the $L/\beta$
enhancement comes from the short thermal cycle
near the rotational axis. The construction of
a globally regular solution and the evaluation
of its complete on-shell action are left for
future work~\cite{MukherjeeWorkInProgress}.


\suppsubheading{Conformally coupled scalar: hyperbolic heat kernel}

We first consider a conformally coupled scalar on
$S^1\times\HH^3$. Using the standard scalar heat
kernel on $\HH^3$~\cite{GiombiMaloneyYin},
the spatial heat trace for a nontrivial rotational
image is
\begin{equation}
 \frac1L\int_{\HH^3}d^3x\,
 K_{\HH^3}(s;x,R_{k\theta}x)e^s
 =\frac1{\sqrt{4\pi s}}\,
 \frac1{4\sin^2(k\theta/2)}.
 \label{eq:sOrbitalScalar}
\end{equation}
Here $R_{k\theta}$ denotes a rotation by
$k\theta$, and the factor $e^s$ cancels the
curvature contribution through the conformal
coupling. Combining this with the thermal
heat kernel on $S^1$ gives the scalar
partition function.

To regulate the winding sum, we first introduce
a transverse cutoff $r\leq r_{\max}$.
The heat-kernel calculation then gives
\begin{equation}
 \frac{\log Z_n^{\rm sc}(\theta;r_{\max})}{L}
 \bigg|_{k\ne0}
 =\frac{1}{8\pi^2n}
 \sum_{k=1}^\infty
 \frac{\Sigma_k^2}
 {k^2s_k^2(B_k^2+\Sigma_k^2)},
 \label{eq:sScalarFiniteCut}
\end{equation}
where
\begin{equation}
 s_k=\left|\sin\frac{k\theta}{2}\right|,
 \qquad
 B_k=\beta k,\qquad
 \Sigma_k=2\operatorname{arsinh}
 (s_k\sinh r_{\max}).
\end{equation}
This expression is finite and absolutely
convergent at every finite cutoff, including
the resonant windings with $s_k=0$.
The transverse cutoff is auxiliary and differs
from the inherited CHM cutoff in
Eq.~\eqref{eq:scut}.

Removing the cutoff at each fixed nonresonant
winding before summing gives the formal series
\begin{equation}
 K_{\scal}^{\rm formal}(\theta)
 =\frac{1}{8\pi^2}
 \sum_{k=1}^\infty
 \frac{1}{k^2\sin^2(k\theta/2)}.
 \label{eq:sScalarFormal}
\end{equation}
This series is not pointwise convergent.
At rational $\theta/(2\pi)$, the exact-return
windings are singular, while at irrational
angles Dirichlet approximation gives infinitely
many summands that do not approach zero.
The finite-cutoff expression regulates these
contributions, which are related to angular
fractals~\cite{AngularFractals}.

We extract the local perturbative coefficients
by expanding in small $\theta$ and using
the Mellin/zeta prescription,
\begin{align}
 K_{\scal}^{\pert}(\theta)
 &=\frac{1}{8\pi^2}
 \left[
 \frac{4\zeta(4)}{\theta^2}
 +\frac{\zeta(2)}{3}
 +\frac{\zeta(0)\theta^2}{60}
 \right]\nonumber\\
 &=\frac{\pi^2}{180\theta^2}
 +\frac1{144}
 -\frac{\theta^2}{960\pi^2}.
 \label{eq:sScalarK}
\end{align}
All higher analytic terms vanish in this
prescription because $\zeta(-2j)=0$ for
positive integers $j$.

\suppsubheading{Direct rotating-sphere scalar character}

A separate operator-counting route gives the same scalar coefficients without
using the hyperbolic heat kernel. The conformally coupled real scalar
one-particle partition function on the unit round $S^3$, with two angular
displacements, is
\begin{equation}
 f_{\scal}(\beta_S,\alpha_1,\alpha_2)
 =\frac{\sinh\beta_S}
 {2(\cosh\beta_S-\cos\alpha_1)
 (\cosh\beta_S-\cos\alpha_2)}.
 \label{eq:sScalarLetter}
\end{equation}
The bosonic multiparticle trace is
$\log Z_{S^3}^{\scal}=\sum_{k=1}^{\infty}k^{-1}
 f_{\scal}(k\beta_S,k\alpha_1,k\alpha_2)$.
In the one-spin specialization $\alpha_1=0$, $\alpha_2=\theta$, taking
the leading $\beta_S\to0$ term for \emph{each fixed winding} gives
\begin{equation}
 \left.\beta_S\log Z_{S^3}^{\scal}(\beta_S,0,\theta)
 \right|_{\mathrm{formal}}=\frac12\sum_{k\ge1}
 \frac{1}{k^2\sin^2(k\theta/2)}.
 \label{eq:sScalarSphere}
\end{equation}
The geometric fixed-circle length is $2\pi$. Combining this length
with $\beta=2\pi n$, the common local coefficient is
$K^{\pert}=\left.\beta_S\log Z_{S^3}\right|_{\rm formal}/(4\pi^2)$.
Thus Eq.~\eqref{eq:sScalarSphere} reproduces
Eq.~\eqref{eq:sScalarFormal} in the same coefficient prescription.

\suppsubheading{Dirac field: hyperbolic and rotating-sphere characters}

For a free Dirac fermion, the spinor character on
$\HH^3$, together with the antiperiodic thermal
boundary condition, gives the following formal
local contribution:
\begin{equation}
 K_{\Dir,H}^{\rm formal}(\theta)
 =\frac1{2\pi^2}\sum_{k\ge1}
 \frac{(-1)^{k+1}\cos(k\theta/2)}
 {k^2\sin^2(k\theta/2)}.
 \label{eq:sDiracFormal}
\end{equation}
Here the cosine factor comes from the spinor
trace under spatial rotations, while
$(-1)^{k+1}$ accounts for fermionic statistics.
The normalization corresponds to four
on-shell Dirac degrees of freedom.

We now obtain the same expression independently
from the rotating-$S^3$ operator character.
For a Dirac fermion, including particle and
antiparticle states, the one-particle character
at $\alpha_1=0$, $\alpha_2=\theta$ is
\begin{equation}
 f_{\Dir}(\beta_S,0,\theta)
 =\frac{8x^{3/2}\cos(\theta/2)}
 {(1-x)(1-2x\cos\theta+x^2)},
 \qquad x=e^{-\beta_S}.
\end{equation}
The corresponding thermal partition function is
\begin{equation}
 \log Z_{S^3}^{\Dir}
 =\sum_{k\ge1}\frac{(-1)^{k+1}}{k}
 f_{\Dir}(k\beta_S,0,k\theta).
\end{equation}
Expanding at small $\beta_S$ for each fixed
winding number $k$, we obtain
\begin{equation}
 \left.\beta_S\log Z_{S^3}^{\Dir}
 \right|_{\rm formal}
 =2\sum_{k\ge1}
 \frac{(-1)^{k+1}\cos(k\theta/2)}
 {k^2\sin^2(k\theta/2)}.
\end{equation}
Thus,
\begin{equation}
 K_{\Dir,H}^{\rm formal}(\theta)
 =\frac1{4\pi^2}
 \left.\beta_S\log Z_{S^3}^{\Dir}
 \right|_{\rm formal}.
 \label{eq:sDiracMatch}
\end{equation}
This establishes the agreement of the
perturbative local coefficients on the two
geometries.

Using
\begin{equation}
 \frac{\cos(x/2)}{\sin^2(x/2)}
 =\frac4{x^2}-\frac16
 -\frac{7x^2}{480}+\cdots,
 \label{eq:sDiracSeries}
\end{equation}
we find
\begin{equation}
 K_{\Dir}^{\pert}(\theta)
 =\frac{7\pi^2}{360\theta^2}
 -\frac1{144}
 -\frac{7\theta^2}{1920\pi^2}.
 \label{eq:sDiracK}
\end{equation}
The leading term reproduces the flat-space
thermal coefficient $c'_{\Dir}=7\pi^2/180$.
All higher even-power terms vanish because
$\eta(-2j)=0$ for $j\ge1$, where
$\eta(s)=(1-2^{1-s})\zeta(s)$.

The equality holds for the formal perturbative
local expansion. The complete fixed-angle
partition functions can contain additional
winding contributions, which are not captured
by this coefficient extraction.


\putbib[references]

\end{bibunit}


\begin{thebibliography}{42}%
\makeatletter
\providecommand \@ifxundefined [1]{%
 \@ifx{#1\undefined}
}%
\providecommand \@ifnum [1]{%
 \ifnum #1\expandafter \@firstoftwo
 \else \expandafter \@secondoftwo
 \fi
}%
\providecommand \@ifx [1]{%
 \ifx #1\expandafter \@firstoftwo
 \else \expandafter \@secondoftwo
 \fi
}%
\providecommand \natexlab [1]{#1}%
\providecommand \enquote  [1]{``#1''}%
\providecommand \bibnamefont  [1]{#1}%
\providecommand \bibfnamefont [1]{#1}%
\providecommand \citenamefont [1]{#1}%
\providecommand \href@noop [0]{\@secondoftwo}%
\providecommand \href [0]{\begingroup \@sanitize@url \@href}%
\providecommand \@href[1]{\@@startlink{#1}\@@href}%
\providecommand \@@href[1]{\endgroup#1\@@endlink}%
\providecommand \@sanitize@url [0]{\catcode `\\12\catcode `\$12\catcode
  `\&12\catcode `\#12\catcode `\^12\catcode `\_12\catcode `\%12\relax}%
\providecommand \@@startlink[1]{}%
\providecommand \@@endlink[0]{}%
\providecommand \url  [0]{\begingroup\@sanitize@url \@url }%
\providecommand \@url [1]{\endgroup\@href {#1}{\urlprefix }}%
\providecommand \urlprefix  [0]{URL }%
\providecommand \Eprint [0]{\href }%
\providecommand \doibase [0]{https://doi.org/}%
\providecommand \selectlanguage [0]{\@gobble}%
\providecommand \bibinfo  [0]{\@secondoftwo}%
\providecommand \bibfield  [0]{\@secondoftwo}%
\providecommand \translation [1]{[#1]}%
\providecommand \BibitemOpen [0]{}%
\providecommand \bibitemStop [0]{}%
\providecommand \bibitemNoStop [0]{.\EOS\space}%
\providecommand \EOS [0]{\spacefactor3000\relax}%
\providecommand \BibitemShut  [1]{\csname bibitem#1\endcsname}%
\let\auto@bib@innerbib\@empty
\bibitem [{\citenamefont {Maldacena}(1998)}]{Maldacena:1997re}%
  \BibitemOpen
  \bibfield  {author} {\bibinfo {author} {\bibfnamefont {J.~M.}\ \bibnamefont
  {Maldacena}},\ }\href {https://doi.org/10.4310/ATMP.1998.v2.n2.a1} {\bibfield
   {journal} {\bibinfo  {journal} {Adv. Theor. Math. Phys.}\ }\textbf {\bibinfo
  {volume} {2}},\ \bibinfo {pages} {231} (\bibinfo {year} {1998})},\ \Eprint
  {https://arxiv.org/abs/hep-th/9711200} {arXiv:hep-th/9711200} \BibitemShut
  {NoStop}%
\bibitem [{\citenamefont {Bhattacharyya}\ \emph {et~al.}(2008)\citenamefont
  {Bhattacharyya}, \citenamefont {Lahiri}, \citenamefont {Loganayagam},\ and\
  \citenamefont {Minwalla}}]{BhattacharyyaRotating}%
  \BibitemOpen
  \bibfield  {author} {\bibinfo {author} {\bibfnamefont {S.}~\bibnamefont
  {Bhattacharyya}}, \bibinfo {author} {\bibfnamefont {S.}~\bibnamefont
  {Lahiri}}, \bibinfo {author} {\bibfnamefont {R.}~\bibnamefont
  {Loganayagam}},\ and\ \bibinfo {author} {\bibfnamefont {S.}~\bibnamefont
  {Minwalla}},\ }\href {https://doi.org/10.1088/1126-6708/2008/09/054}
  {\bibfield  {journal} {\bibinfo  {journal} {JHEP}\ }\textbf {\bibinfo
  {volume} {09}}\bibfield  {number} {\bibinfo  {number} { (2008)},\ \bibinfo
  {pages} {054}},\ }\Eprint {https://arxiv.org/abs/0708.1770} {arXiv:0708.1770
  [hep-th]} \BibitemShut {NoStop}%
\bibitem [{\citenamefont {Banerjee}\ \emph {et~al.}(2012)\citenamefont
  {Banerjee}, \citenamefont {Bhattacharya}, \citenamefont {Bhattacharyya},
  \citenamefont {Jain}, \citenamefont {Minwalla},\ and\ \citenamefont
  {Sharma}}]{BanerjeeEquilibrium}%
  \BibitemOpen
  \bibfield  {author} {\bibinfo {author} {\bibfnamefont {N.}~\bibnamefont
  {Banerjee}}, \bibinfo {author} {\bibfnamefont {J.}~\bibnamefont
  {Bhattacharya}}, \bibinfo {author} {\bibfnamefont {S.}~\bibnamefont
  {Bhattacharyya}}, \bibinfo {author} {\bibfnamefont {S.}~\bibnamefont {Jain}},
  \bibinfo {author} {\bibfnamefont {S.}~\bibnamefont {Minwalla}},\ and\
  \bibinfo {author} {\bibfnamefont {T.}~\bibnamefont {Sharma}},\ }\href
  {https://doi.org/10.1007/JHEP09(2012)046} {\bibfield  {journal} {\bibinfo
  {journal} {JHEP}\ }\textbf {\bibinfo {volume} {09}}\bibfield  {number}
  {\bibinfo  {number} { (2012)},\ \bibinfo {pages} {046}},\ }\Eprint
  {https://arxiv.org/abs/1203.3544} {arXiv:1203.3544 [hep-th]} \BibitemShut
  {NoStop}%
\bibitem [{\citenamefont {Jensen}\ \emph {et~al.}(2012)\citenamefont {Jensen},
  \citenamefont {Kaminski}, \citenamefont {Kovtun}, \citenamefont {Meyer},
  \citenamefont {Ritz},\ and\ \citenamefont {Yarom}}]{JensenEquilibrium}%
  \BibitemOpen
  \bibfield  {author} {\bibinfo {author} {\bibfnamefont {K.}~\bibnamefont
  {Jensen}}, \bibinfo {author} {\bibfnamefont {M.}~\bibnamefont {Kaminski}},
  \bibinfo {author} {\bibfnamefont {P.}~\bibnamefont {Kovtun}}, \bibinfo
  {author} {\bibfnamefont {R.}~\bibnamefont {Meyer}}, \bibinfo {author}
  {\bibfnamefont {A.}~\bibnamefont {Ritz}},\ and\ \bibinfo {author}
  {\bibfnamefont {A.}~\bibnamefont {Yarom}},\ }\href
  {https://doi.org/10.1103/PhysRevLett.109.101601} {\bibfield  {journal}
  {\bibinfo  {journal} {Phys. Rev. Lett.}\ }\textbf {\bibinfo {volume} {109}},\
  \bibinfo {pages} {101601} (\bibinfo {year} {2012})},\ \Eprint
  {https://arxiv.org/abs/1203.3556} {arXiv:1203.3556 [hep-th]} \BibitemShut
  {NoStop}%
\bibitem [{\citenamefont {Benjamin}\ \emph
  {et~al.}(2024{\natexlab{a}})\citenamefont {Benjamin}, \citenamefont {Lee},
  \citenamefont {Ooguri},\ and\ \citenamefont
  {Simmons-Duffin}}]{BenjaminOoguri}%
  \BibitemOpen
  \bibfield  {author} {\bibinfo {author} {\bibfnamefont {N.}~\bibnamefont
  {Benjamin}}, \bibinfo {author} {\bibfnamefont {J.}~\bibnamefont {Lee}},
  \bibinfo {author} {\bibfnamefont {H.}~\bibnamefont {Ooguri}},\ and\ \bibinfo
  {author} {\bibfnamefont {D.}~\bibnamefont {Simmons-Duffin}},\ }\href
  {https://doi.org/10.1007/JHEP03(2024)115} {\bibfield  {journal} {\bibinfo
  {journal} {JHEP}\ }\textbf {\bibinfo {volume} {03}}\bibfield  {number}
  {\bibinfo  {number} { (2024)},\ \bibinfo {pages} {115}},\ }\Eprint
  {https://arxiv.org/abs/2306.08031} {arXiv:2306.08031 [hep-th]} \BibitemShut
  {NoStop}%
\bibitem [{\citenamefont {Anand}\ \emph {et~al.}(2026)\citenamefont {Anand},
  \citenamefont {Benjamin}, \citenamefont {Kumar}, \citenamefont {Minwalla},
  \citenamefont {Mukherjee}, \citenamefont {Pal},\ and\ \citenamefont
  {Rahaman}}]{AnandSemiUniversality}%
  \BibitemOpen
  \bibfield  {author} {\bibinfo {author} {\bibfnamefont {H.}~\bibnamefont
  {Anand}}, \bibinfo {author} {\bibfnamefont {N.}~\bibnamefont {Benjamin}},
  \bibinfo {author} {\bibfnamefont {V.}~\bibnamefont {Kumar}}, \bibinfo
  {author} {\bibfnamefont {S.}~\bibnamefont {Minwalla}}, \bibinfo {author}
  {\bibfnamefont {J.}~\bibnamefont {Mukherjee}}, \bibinfo {author}
  {\bibfnamefont {S.}~\bibnamefont {Pal}},\ and\ \bibinfo {author}
  {\bibfnamefont {A.}~\bibnamefont {Rahaman}},\ }\href
  {https://doi.org/10.21468/SciPostPhys.21.2.042} {\bibfield  {journal}
  {\bibinfo  {journal} {SciPost Phys.}\ }\textbf {\bibinfo {volume} {21}},\
  \bibinfo {pages} {042} (\bibinfo {year} {2026})},\ \Eprint
  {https://arxiv.org/abs/2512.00158} {arXiv:2512.00158 [hep-th]} \BibitemShut
  {NoStop}%
\bibitem [{\citenamefont {Komargodski}\ \emph {et~al.}(2026)\citenamefont
  {Komargodski}, \citenamefont {Miscioscia},\ and\ \citenamefont
  {Popov}}]{KomargodskiRegge}%
  \BibitemOpen
  \bibfield  {author} {\bibinfo {author} {\bibfnamefont {Z.}~\bibnamefont
  {Komargodski}}, \bibinfo {author} {\bibfnamefont {A.}~\bibnamefont
  {Miscioscia}},\ and\ \bibinfo {author} {\bibfnamefont {F.~K.}\ \bibnamefont
  {Popov}},\ }\href@noop {} {\bibinfo {title} {Regge's inferno}} (\bibinfo
  {year} {2026}),\ \Eprint {https://arxiv.org/abs/2603.10197} {arXiv:2603.10197
  [hep-th]} \BibitemShut {NoStop}%
\bibitem [{\citenamefont {Advant}\ \emph
  {et~al.}(2026{\natexlab{a}})\citenamefont {Advant}, \citenamefont {Anand},
  \citenamefont {Benjamin}, \citenamefont {Kumar}, \citenamefont {Minwalla},
  \citenamefont {Mukherjee}, \citenamefont {Pal}, \citenamefont {Rahaman},\
  and\ \citenamefont {Ray}}]{FromZtoa}%
  \BibitemOpen
  \bibfield  {author} {\bibinfo {author} {\bibfnamefont {A.}~\bibnamefont
  {Advant}}, \bibinfo {author} {\bibfnamefont {H.}~\bibnamefont {Anand}},
  \bibinfo {author} {\bibfnamefont {N.}~\bibnamefont {Benjamin}}, \bibinfo
  {author} {\bibfnamefont {V.}~\bibnamefont {Kumar}}, \bibinfo {author}
  {\bibfnamefont {S.}~\bibnamefont {Minwalla}}, \bibinfo {author}
  {\bibfnamefont {J.}~\bibnamefont {Mukherjee}}, \bibinfo {author}
  {\bibfnamefont {S.}~\bibnamefont {Pal}}, \bibinfo {author} {\bibfnamefont
  {A.}~\bibnamefont {Rahaman}},\ and\ \bibinfo {author} {\bibfnamefont
  {P.}~\bibnamefont {Ray}},\ }\href@noop {} {\bibinfo {title} {From {$Z$} to
  {$a$}: High-temperature relations, subleading semi-universality, and
  conformal anomalies}} (\bibinfo {year} {2026}{\natexlab{a}}),\ \Eprint
  {https://arxiv.org/abs/2609.03013} {arXiv:2609.03013 [hep-th]} \BibitemShut
  {NoStop}%
\bibitem [{\citenamefont {Kim}\ \emph {et~al.}(2023)\citenamefont {Kim},
  \citenamefont {Kundu}, \citenamefont {Lee}, \citenamefont {Lee},
  \citenamefont {Minwalla},\ and\ \citenamefont {Patel}}]{Kim:2023sig}%
  \BibitemOpen
  \bibfield  {author} {\bibinfo {author} {\bibfnamefont {S.}~\bibnamefont
  {Kim}}, \bibinfo {author} {\bibfnamefont {S.}~\bibnamefont {Kundu}}, \bibinfo
  {author} {\bibfnamefont {E.}~\bibnamefont {Lee}}, \bibinfo {author}
  {\bibfnamefont {J.}~\bibnamefont {Lee}}, \bibinfo {author} {\bibfnamefont
  {S.}~\bibnamefont {Minwalla}},\ and\ \bibinfo {author} {\bibfnamefont
  {C.}~\bibnamefont {Patel}},\ }\href {https://doi.org/10.1007/JHEP11(2023)024}
  {\bibfield  {journal} {\bibinfo  {journal} {JHEP}\ }\textbf {\bibinfo
  {volume} {11}},\ \bibinfo {pages} {024}},\ \Eprint
  {https://arxiv.org/abs/2305.08922} {arXiv:2305.08922 [hep-th]} \BibitemShut
  {NoStop}%
\bibitem [{\citenamefont {Bajaj}\ \emph {et~al.}(2026)\citenamefont {Bajaj},
  \citenamefont {Kumar}, \citenamefont {Minwalla}, \citenamefont {Mukherjee},\
  and\ \citenamefont {Rahaman}}]{Bajaj:2024utv}%
  \BibitemOpen
  \bibfield  {author} {\bibinfo {author} {\bibfnamefont {K.}~\bibnamefont
  {Bajaj}}, \bibinfo {author} {\bibfnamefont {V.}~\bibnamefont {Kumar}},
  \bibinfo {author} {\bibfnamefont {S.}~\bibnamefont {Minwalla}}, \bibinfo
  {author} {\bibfnamefont {J.}~\bibnamefont {Mukherjee}},\ and\ \bibinfo
  {author} {\bibfnamefont {A.}~\bibnamefont {Rahaman}},\ }\href
  {https://doi.org/10.21468/SciPostPhys.20.6.178} {\bibfield  {journal}
  {\bibinfo  {journal} {SciPost Phys.}\ }\textbf {\bibinfo {volume} {20}},\
  \bibinfo {pages} {178} (\bibinfo {year} {2026})},\ \Eprint
  {https://arxiv.org/abs/2412.06904} {arXiv:2412.06904 [hep-th]} \BibitemShut
  {NoStop}%
\bibitem [{\citenamefont {Lee}(2026)}]{Lee:2026azy}%
  \BibitemOpen
  \bibfield  {author} {\bibinfo {author} {\bibfnamefont {E.}~\bibnamefont
  {Lee}},\ }\href@noop {} {\bibinfo {title} {Equilibrium partition function of
  non-relativistic cfts in harmonic trap}} (\bibinfo {year} {2026}),\ \Eprint
  {https://arxiv.org/abs/2603.09856} {arXiv:2603.09856 [hep-th]} \BibitemShut
  {NoStop}%
\bibitem [{\citenamefont {Mukherjee}\ and\ \citenamefont
  {Ray}(2026)}]{Mukherjee:2026dfu}%
  \BibitemOpen
  \bibfield  {author} {\bibinfo {author} {\bibfnamefont {J.}~\bibnamefont
  {Mukherjee}}\ and\ \bibinfo {author} {\bibfnamefont {P.}~\bibnamefont
  {Ray}},\ }\href {https://doi.org/10.1007/JHEP08(2026)182} {\bibfield
  {journal} {\bibinfo  {journal} {JHEP}\ }\textbf {\bibinfo {volume} {08}},\
  \bibinfo {pages} {182}},\ \Eprint {https://arxiv.org/abs/2606.08461}
  {arXiv:2606.08461 [hep-th]} \BibitemShut {NoStop}%
\bibitem [{\citenamefont {Mishra}(2026)}]{Mishra:2026bpi}%
  \BibitemOpen
  \bibfield  {author} {\bibinfo {author} {\bibfnamefont {V.}~\bibnamefont
  {Mishra}},\ }\href@noop {} {\bibinfo {title} {{Semi-Universality of $O(N)$
  model on $S^1\times S^2$}}} (\bibinfo {year} {2026}),\ \Eprint
  {https://arxiv.org/abs/2609.25622} {arXiv:2609.25622 [hep-th]} \BibitemShut
  {NoStop}%
\bibitem [{\citenamefont {David}\ and\ \citenamefont
  {Kumar}(2026)}]{David:2026yis}%
  \BibitemOpen
  \bibfield  {author} {\bibinfo {author} {\bibfnamefont {J.~R.}\ \bibnamefont
  {David}}\ and\ \bibinfo {author} {\bibfnamefont {S.}~\bibnamefont {Kumar}},\
  }\href@noop {} {\bibinfo {title} {{The large $N$ vector model with angular
  velocity}}} (\bibinfo {year} {2026}),\ \Eprint
  {https://arxiv.org/abs/2608.31151} {arXiv:2608.31151 [hep-th]} \BibitemShut
  {NoStop}%
\bibitem [{\citenamefont {Miscioscia}\ \emph {et~al.}(2026)\citenamefont
  {Miscioscia}, \citenamefont {Pavshinkin},\ and\ \citenamefont
  {Popov}}]{Miscioscia:2026bvl}%
  \BibitemOpen
  \bibfield  {author} {\bibinfo {author} {\bibfnamefont {A.}~\bibnamefont
  {Miscioscia}}, \bibinfo {author} {\bibfnamefont {D.}~\bibnamefont
  {Pavshinkin}},\ and\ \bibinfo {author} {\bibfnamefont {F.~K.}\ \bibnamefont
  {Popov}},\ }\href@noop {} {\bibinfo {title} {{Regge in Wonderland}}}
  (\bibinfo {year} {2026}),\ \Eprint {https://arxiv.org/abs/2609.35455}
  {arXiv:2609.35455 [hep-th]} \BibitemShut {NoStop}%
\bibitem [{\citenamefont {Casini}\ \emph {et~al.}(2011)\citenamefont {Casini},
  \citenamefont {Huerta},\ and\ \citenamefont {Myers}}]{CHM}%
  \BibitemOpen
  \bibfield  {author} {\bibinfo {author} {\bibfnamefont {H.}~\bibnamefont
  {Casini}}, \bibinfo {author} {\bibfnamefont {M.}~\bibnamefont {Huerta}},\
  and\ \bibinfo {author} {\bibfnamefont {R.~C.}\ \bibnamefont {Myers}},\ }\href
  {https://doi.org/10.1007/JHEP05(2011)036} {\bibfield  {journal} {\bibinfo
  {journal} {JHEP}\ }\textbf {\bibinfo {volume} {05}}\bibfield  {number}
  {\bibinfo  {number} { (2011)},\ \bibinfo {pages} {036}},\ }\Eprint
  {https://arxiv.org/abs/1102.0440} {arXiv:1102.0440 [hep-th]} \BibitemShut
  {NoStop}%
\bibitem [{\citenamefont {Casini}\ and\ \citenamefont
  {Huerta}(2010)}]{CasiniHuertaSphere}%
  \BibitemOpen
  \bibfield  {author} {\bibinfo {author} {\bibfnamefont {H.}~\bibnamefont
  {Casini}}\ and\ \bibinfo {author} {\bibfnamefont {M.}~\bibnamefont
  {Huerta}},\ }\href {https://doi.org/10.1016/j.physletb.2010.09.054}
  {\bibfield  {journal} {\bibinfo  {journal} {Phys. Lett. B}\ }\textbf
  {\bibinfo {volume} {694}},\ \bibinfo {pages} {167} (\bibinfo {year}
  {2010})},\ \Eprint {https://arxiv.org/abs/1007.1813} {arXiv:1007.1813
  [hep-th]} \BibitemShut {NoStop}%
\bibitem [{\citenamefont {Klebanov}\ \emph {et~al.}(2012)\citenamefont
  {Klebanov}, \citenamefont {Pufu}, \citenamefont {Sachdev},\ and\
  \citenamefont {Safdi}}]{KlebanovFreeRenyi}%
  \BibitemOpen
  \bibfield  {author} {\bibinfo {author} {\bibfnamefont {I.~R.}\ \bibnamefont
  {Klebanov}}, \bibinfo {author} {\bibfnamefont {S.~S.}\ \bibnamefont {Pufu}},
  \bibinfo {author} {\bibfnamefont {S.}~\bibnamefont {Sachdev}},\ and\ \bibinfo
  {author} {\bibfnamefont {B.~R.}\ \bibnamefont {Safdi}},\ }\href
  {https://doi.org/10.1007/JHEP04(2012)074} {\bibfield  {journal} {\bibinfo
  {journal} {JHEP}\ }\textbf {\bibinfo {volume} {04}}\bibfield  {number}
  {\bibinfo  {number} { (2012)},\ \bibinfo {pages} {074}},\ }\Eprint
  {https://arxiv.org/abs/1111.6290} {arXiv:1111.6290 [hep-th]} \BibitemShut
  {NoStop}%
\bibitem [{\citenamefont {Hung}\ \emph {et~al.}(2011)\citenamefont {Hung},
  \citenamefont {Myers}, \citenamefont {Smolkin},\ and\ \citenamefont
  {Yale}}]{HungMyersSmolkinYale}%
  \BibitemOpen
  \bibfield  {author} {\bibinfo {author} {\bibfnamefont {L.-Y.}\ \bibnamefont
  {Hung}}, \bibinfo {author} {\bibfnamefont {R.~C.}\ \bibnamefont {Myers}},
  \bibinfo {author} {\bibfnamefont {M.}~\bibnamefont {Smolkin}},\ and\ \bibinfo
  {author} {\bibfnamefont {A.}~\bibnamefont {Yale}},\ }\href
  {https://doi.org/10.1007/JHEP12(2011)047} {\bibfield  {journal} {\bibinfo
  {journal} {JHEP}\ }\textbf {\bibinfo {volume} {12}}\bibfield  {number}
  {\bibinfo  {number} { (2011)},\ \bibinfo {pages} {047}},\ }\Eprint
  {https://arxiv.org/abs/1110.1084} {arXiv:1110.1084 [hep-th]} \BibitemShut
  {NoStop}%
\bibitem [{\citenamefont {David}\ and\ \citenamefont
  {Mukherjee}(2021)}]{DavidMukherjee2021}%
  \BibitemOpen
  \bibfield  {author} {\bibinfo {author} {\bibfnamefont {J.~R.}\ \bibnamefont
  {David}}\ and\ \bibinfo {author} {\bibfnamefont {J.}~\bibnamefont
  {Mukherjee}},\ }\href {https://doi.org/10.1007/JHEP01(2021)202} {\bibfield
  {journal} {\bibinfo  {journal} {JHEP}\ }\textbf {\bibinfo {volume}
  {01}}\bibfield  {number} {\bibinfo  {number} { (2021)},\ \bibinfo {pages}
  {202}},\ }\Eprint {https://arxiv.org/abs/2005.08402} {arXiv:2005.08402
  [hep-th]} \BibitemShut {NoStop}%
\bibitem [{\citenamefont {Mukherjee}(2021)}]{MukherjeeHigherDerivative}%
  \BibitemOpen
  \bibfield  {author} {\bibinfo {author} {\bibfnamefont {J.}~\bibnamefont
  {Mukherjee}},\ }\href {https://doi.org/10.1007/JHEP10(2021)236} {\bibfield
  {journal} {\bibinfo  {journal} {JHEP}\ }\textbf {\bibinfo {volume}
  {10}}\bibfield  {number} {\bibinfo  {number} { (2021)},\ \bibinfo {pages}
  {236}},\ }\Eprint {https://arxiv.org/abs/2108.00929} {arXiv:2108.00929
  [hep-th]} \BibitemShut {NoStop}%
\bibitem [{\citenamefont {Mukherjee}(2022)}]{MukherjeeHigherSpin}%
  \BibitemOpen
  \bibfield  {author} {\bibinfo {author} {\bibfnamefont {J.}~\bibnamefont
  {Mukherjee}},\ }\href {https://doi.org/10.1007/JHEP04(2022)071} {\bibfield
  {journal} {\bibinfo  {journal} {JHEP}\ }\textbf {\bibinfo {volume}
  {04}}\bibfield  {number} {\bibinfo  {number} { (2022)},\ \bibinfo {pages}
  {071}},\ }\Eprint {https://arxiv.org/abs/2112.15461} {arXiv:2112.15461
  [hep-th]} \BibitemShut {NoStop}%
\bibitem [{\citenamefont {Kusuki}\ \emph {et~al.}(2025)\citenamefont {Kusuki},
  \citenamefont {Ooguri},\ and\ \citenamefont {Pal}}]{KusukiOoguriPal}%
  \BibitemOpen
  \bibfield  {author} {\bibinfo {author} {\bibfnamefont {Y.}~\bibnamefont
  {Kusuki}}, \bibinfo {author} {\bibfnamefont {H.}~\bibnamefont {Ooguri}},\
  and\ \bibinfo {author} {\bibfnamefont {S.}~\bibnamefont {Pal}},\ }\href
  {https://doi.org/10.1103/fsg7-bs7q} {\bibfield  {journal} {\bibinfo
  {journal} {Phys. Rev. Lett.}\ }\textbf {\bibinfo {volume} {135}},\ \bibinfo
  {pages} {061603} (\bibinfo {year} {2025})},\ \Eprint
  {https://arxiv.org/abs/2503.24353} {arXiv:2503.24353 [hep-th]} \BibitemShut
  {NoStop}%
\bibitem [{\citenamefont {Advant}\ \emph
  {et~al.}(2026{\natexlab{b}})\citenamefont {Advant}, \citenamefont {Anand},
  \citenamefont {Benjamin}, \citenamefont {Gadde}, \citenamefont {Kumar},
  \citenamefont {Minwalla}, \citenamefont {Mukherjee}, \citenamefont {Pal},
  \citenamefont {Rahaman},\ and\ \citenamefont {Ray}}]{Advant:2026zft}%
  \BibitemOpen
  \bibfield  {author} {\bibinfo {author} {\bibfnamefont {A.}~\bibnamefont
  {Advant}}, \bibinfo {author} {\bibfnamefont {H.}~\bibnamefont {Anand}},
  \bibinfo {author} {\bibfnamefont {N.}~\bibnamefont {Benjamin}}, \bibinfo
  {author} {\bibfnamefont {A.}~\bibnamefont {Gadde}}, \bibinfo {author}
  {\bibfnamefont {V.}~\bibnamefont {Kumar}}, \bibinfo {author} {\bibfnamefont
  {S.}~\bibnamefont {Minwalla}}, \bibinfo {author} {\bibfnamefont
  {J.}~\bibnamefont {Mukherjee}}, \bibinfo {author} {\bibfnamefont
  {S.}~\bibnamefont {Pal}}, \bibinfo {author} {\bibfnamefont {A.}~\bibnamefont
  {Rahaman}},\ and\ \bibinfo {author} {\bibfnamefont {P.}~\bibnamefont {Ray}},\
  }\href@noop {} {\bibinfo {title} {{From Z to all: cyclic sum rules for
  thermal CFT partition functions}}} (\bibinfo {year} {2026}{\natexlab{b}}),\
  \Eprint {https://arxiv.org/abs/2611.05281} {arXiv:2611.05281 [hep-th]}
  \BibitemShut {NoStop}%
\bibitem [{\citenamefont {Dehghani}(2002)}]{Dehghani}%
  \BibitemOpen
  \bibfield  {author} {\bibinfo {author} {\bibfnamefont {M.~H.}\ \bibnamefont
  {Dehghani}},\ }\href {https://doi.org/10.1103/PhysRevD.65.124002} {\bibfield
  {journal} {\bibinfo  {journal} {Phys. Rev. D}\ }\textbf {\bibinfo {volume}
  {65}},\ \bibinfo {pages} {124002} (\bibinfo {year} {2002})},\ \Eprint
  {https://arxiv.org/abs/hep-th/0203049} {arXiv:hep-th/0203049} \BibitemShut
  {NoStop}%
\bibitem [{\citenamefont {He}(2022)}]{HeRotating}%
  \BibitemOpen
  \bibfield  {author} {\bibinfo {author} {\bibfnamefont {B.}~\bibnamefont
  {He}},\ }\href@noop {} {\bibinfo {title} {Holographic calculations of
  {R\'enyi} entropy from rotating topological black holes}} (\bibinfo {year}
  {2022}),\ \Eprint {https://arxiv.org/abs/2207.05354} {arXiv:2207.05354
  [hep-th]} \BibitemShut {NoStop}%
\bibitem [{\citenamefont {Balasubramanian}\ and\ \citenamefont
  {Kraus}(1999)}]{BalasubramanianKraus}%
  \BibitemOpen
  \bibfield  {author} {\bibinfo {author} {\bibfnamefont {V.}~\bibnamefont
  {Balasubramanian}}\ and\ \bibinfo {author} {\bibfnamefont {P.}~\bibnamefont
  {Kraus}},\ }\href {https://doi.org/10.1007/s002200050764} {\bibfield
  {journal} {\bibinfo  {journal} {Commun. Math. Phys.}\ }\textbf {\bibinfo
  {volume} {208}},\ \bibinfo {pages} {413} (\bibinfo {year} {1999})},\ \Eprint
  {https://arxiv.org/abs/hep-th/9902121} {arXiv:hep-th/9902121} \BibitemShut
  {NoStop}%
\bibitem [{\citenamefont {Skenderis}(2002)}]{Skenderis}%
  \BibitemOpen
  \bibfield  {author} {\bibinfo {author} {\bibfnamefont {K.}~\bibnamefont
  {Skenderis}},\ }\href {https://doi.org/10.1088/0264-9381/19/22/306}
  {\bibfield  {journal} {\bibinfo  {journal} {Class. Quant. Grav.}\ }\textbf
  {\bibinfo {volume} {19}},\ \bibinfo {pages} {5849} (\bibinfo {year}
  {2002})},\ \Eprint {https://arxiv.org/abs/hep-th/0209067}
  {arXiv:hep-th/0209067} \BibitemShut {NoStop}%
\bibitem [{\citenamefont {Henningson}\ and\ \citenamefont
  {Skenderis}(1998)}]{HenningsonSkenderis}%
  \BibitemOpen
  \bibfield  {author} {\bibinfo {author} {\bibfnamefont {M.}~\bibnamefont
  {Henningson}}\ and\ \bibinfo {author} {\bibfnamefont {K.}~\bibnamefont
  {Skenderis}},\ }\href {https://doi.org/10.1088/1126-6708/1998/07/023}
  {\bibfield  {journal} {\bibinfo  {journal} {JHEP}\ }\textbf {\bibinfo
  {volume} {07}}\bibfield  {number} {\bibinfo  {number} { (1998)},\ \bibinfo
  {pages} {023}},\ }\Eprint {https://arxiv.org/abs/hep-th/9806087}
  {arXiv:hep-th/9806087} \BibitemShut {NoStop}%
\bibitem [{\citenamefont {Gibbons}\ \emph {et~al.}(2005)\citenamefont
  {Gibbons}, \citenamefont {Perry},\ and\ \citenamefont
  {Pope}}]{GibbonsPerryPope}%
  \BibitemOpen
  \bibfield  {author} {\bibinfo {author} {\bibfnamefont {G.~W.}\ \bibnamefont
  {Gibbons}}, \bibinfo {author} {\bibfnamefont {M.~J.}\ \bibnamefont {Perry}},\
  and\ \bibinfo {author} {\bibfnamefont {C.~N.}\ \bibnamefont {Pope}},\ }\href
  {https://doi.org/10.1088/0264-9381/22/9/002} {\bibfield  {journal} {\bibinfo
  {journal} {Class. Quant. Grav.}\ }\textbf {\bibinfo {volume} {22}},\ \bibinfo
  {pages} {1503} (\bibinfo {year} {2005})},\ \Eprint
  {https://arxiv.org/abs/hep-th/0408217} {arXiv:hep-th/0408217} \BibitemShut
  {NoStop}%
\bibitem [{Sup(2026)}]{Supplement}%
  \BibitemOpen
  \href@noop {} {} (\bibinfo {year} {2026}),\ \bibinfo {note} {see Supplemental
  Material for the gravitational renormalization and Weyl cocycle, free-field
  image and character computations, conformal-frame conventions, and regulator
  details.}\BibitemShut {Stop}%
\bibitem [{\citenamefont {Donnelly}\ and\ \citenamefont
  {Wall}(2016)}]{DonnellyWall}%
  \BibitemOpen
  \bibfield  {author} {\bibinfo {author} {\bibfnamefont {W.}~\bibnamefont
  {Donnelly}}\ and\ \bibinfo {author} {\bibfnamefont {A.~C.}\ \bibnamefont
  {Wall}},\ }\href {https://doi.org/10.1103/PhysRevD.94.104053} {\bibfield
  {journal} {\bibinfo  {journal} {Phys. Rev. D}\ }\textbf {\bibinfo {volume}
  {94}},\ \bibinfo {pages} {104053} (\bibinfo {year} {2016})},\ \Eprint
  {https://arxiv.org/abs/1506.05792} {arXiv:1506.05792 [hep-th]} \BibitemShut
  {NoStop}%
\bibitem [{\citenamefont {Soni}\ and\ \citenamefont
  {Trivedi}(2016)}]{SoniTrivedi2016}%
  \BibitemOpen
  \bibfield  {author} {\bibinfo {author} {\bibfnamefont {R.~M.}\ \bibnamefont
  {Soni}}\ and\ \bibinfo {author} {\bibfnamefont {S.~P.}\ \bibnamefont
  {Trivedi}},\ }\href {https://doi.org/10.1007/JHEP01(2016)136} {\bibfield
  {journal} {\bibinfo  {journal} {JHEP}\ }\textbf {\bibinfo {volume} {01}},\
  \bibinfo {pages} {136}},\ \Eprint {https://arxiv.org/abs/1510.07455}
  {arXiv:1510.07455 [hep-th]} \BibitemShut {NoStop}%
\bibitem [{\citenamefont {David}\ and\ \citenamefont
  {Mukherjee}(2022)}]{DavidMukherjee}%
  \BibitemOpen
  \bibfield  {author} {\bibinfo {author} {\bibfnamefont {J.~R.}\ \bibnamefont
  {David}}\ and\ \bibinfo {author} {\bibfnamefont {J.}~\bibnamefont
  {Mukherjee}},\ }\href {https://doi.org/10.1007/JHEP08(2022)065} {\bibfield
  {journal} {\bibinfo  {journal} {JHEP}\ }\textbf {\bibinfo {volume}
  {08}}\bibfield  {number} {\bibinfo  {number} { (2022)},\ \bibinfo {pages}
  {065}},\ }\Eprint {https://arxiv.org/abs/2201.06043} {arXiv:2201.06043
  [hep-th]} \BibitemShut {NoStop}%
\bibitem [{\citenamefont {Belin}\ \emph {et~al.}(2013)\citenamefont {Belin},
  \citenamefont {Hung}, \citenamefont {Maloney}, \citenamefont {Matsuura},
  \citenamefont {Myers},\ and\ \citenamefont {Sierens}}]{BelinCharged}%
  \BibitemOpen
  \bibfield  {author} {\bibinfo {author} {\bibfnamefont {A.}~\bibnamefont
  {Belin}}, \bibinfo {author} {\bibfnamefont {L.-Y.}\ \bibnamefont {Hung}},
  \bibinfo {author} {\bibfnamefont {A.}~\bibnamefont {Maloney}}, \bibinfo
  {author} {\bibfnamefont {S.}~\bibnamefont {Matsuura}}, \bibinfo {author}
  {\bibfnamefont {R.~C.}\ \bibnamefont {Myers}},\ and\ \bibinfo {author}
  {\bibfnamefont {T.}~\bibnamefont {Sierens}},\ }\href
  {https://doi.org/10.1007/JHEP12(2013)059} {\bibfield  {journal} {\bibinfo
  {journal} {JHEP}\ }\textbf {\bibinfo {volume} {12}}\bibfield  {number}
  {\bibinfo  {number} { (2013)},\ \bibinfo {pages} {059}},\ }\Eprint
  {https://arxiv.org/abs/1310.4180} {arXiv:1310.4180 [hep-th]} \BibitemShut
  {NoStop}%
\bibitem [{\citenamefont {Goldstein}\ and\ \citenamefont
  {Sela}(2018)}]{GoldsteinSela}%
  \BibitemOpen
  \bibfield  {author} {\bibinfo {author} {\bibfnamefont {M.}~\bibnamefont
  {Goldstein}}\ and\ \bibinfo {author} {\bibfnamefont {E.}~\bibnamefont
  {Sela}},\ }\href {https://doi.org/10.1103/PhysRevLett.120.200602} {\bibfield
  {journal} {\bibinfo  {journal} {Phys. Rev. Lett.}\ }\textbf {\bibinfo
  {volume} {120}},\ \bibinfo {pages} {200602} (\bibinfo {year} {2018})},\
  \Eprint {https://arxiv.org/abs/1711.09418} {arXiv:1711.09418} \BibitemShut
  {NoStop}%
\bibitem [{\citenamefont {Benini}\ \emph {et~al.}(2025)\citenamefont {Benini},
  \citenamefont {Godet},\ and\ \citenamefont {Singh}}]{BeniniAsymmetry}%
  \BibitemOpen
  \bibfield  {author} {\bibinfo {author} {\bibfnamefont {F.}~\bibnamefont
  {Benini}}, \bibinfo {author} {\bibfnamefont {V.}~\bibnamefont {Godet}},\ and\
  \bibinfo {author} {\bibfnamefont {A.~H.}\ \bibnamefont {Singh}},\ }\href
  {https://doi.org/10.1093/ptep/ptaf080} {\bibfield  {journal} {\bibinfo
  {journal} {Prog. Theor. Exp. Phys.}\ }\textbf {\bibinfo {volume} {2025}},\
  \bibinfo {pages} {063B05} (\bibinfo {year} {2025})},\ \Eprint
  {https://arxiv.org/abs/2407.07969} {arXiv:2407.07969 [hep-th]} \BibitemShut
  {NoStop}%
\bibitem [{\citenamefont {Benjamin}\ \emph
  {et~al.}(2024{\natexlab{b}})\citenamefont {Benjamin}, \citenamefont {Lee},
  \citenamefont {Pal}, \citenamefont {Simmons-Duffin},\ and\ \citenamefont
  {Xu}}]{AngularFractals}%
  \BibitemOpen
  \bibfield  {author} {\bibinfo {author} {\bibfnamefont {N.}~\bibnamefont
  {Benjamin}}, \bibinfo {author} {\bibfnamefont {J.}~\bibnamefont {Lee}},
  \bibinfo {author} {\bibfnamefont {S.}~\bibnamefont {Pal}}, \bibinfo {author}
  {\bibfnamefont {D.}~\bibnamefont {Simmons-Duffin}},\ and\ \bibinfo {author}
  {\bibfnamefont {Y.}~\bibnamefont {Xu}},\ }\href
  {https://doi.org/10.1007/JHEP11(2024)134} {\bibfield  {journal} {\bibinfo
  {journal} {JHEP}\ }\textbf {\bibinfo {volume} {11}}\bibfield  {number}
  {\bibinfo  {number} { (2024)},\ \bibinfo {pages} {134}},\ }\Eprint
  {https://arxiv.org/abs/2405.17562} {arXiv:2405.17562 [hep-th]} \BibitemShut
  {NoStop}%
\bibitem [{\citenamefont {Deser}\ and\ \citenamefont
  {Schwimmer}(1993)}]{DeserSchwimmer}%
  \BibitemOpen
  \bibfield  {author} {\bibinfo {author} {\bibfnamefont {S.}~\bibnamefont
  {Deser}}\ and\ \bibinfo {author} {\bibfnamefont {A.}~\bibnamefont
  {Schwimmer}},\ }\href {https://doi.org/10.1016/0370-2693(93)90934-A}
  {\bibfield  {journal} {\bibinfo  {journal} {Phys. Lett. B}\ }\textbf
  {\bibinfo {volume} {309}},\ \bibinfo {pages} {279} (\bibinfo {year}
  {1993})},\ \Eprint {https://arxiv.org/abs/hep-th/9302047}
  {arXiv:hep-th/9302047} \BibitemShut {NoStop}%
\bibitem [{\citenamefont {Giombi}\ \emph {et~al.}(2008)\citenamefont {Giombi},
  \citenamefont {Maloney},\ and\ \citenamefont {Yin}}]{GiombiMaloneyYin}%
  \BibitemOpen
  \bibfield  {author} {\bibinfo {author} {\bibfnamefont {S.}~\bibnamefont
  {Giombi}}, \bibinfo {author} {\bibfnamefont {A.}~\bibnamefont {Maloney}},\
  and\ \bibinfo {author} {\bibfnamefont {X.}~\bibnamefont {Yin}},\ }\href
  {https://doi.org/10.1088/1126-6708/2008/08/007} {\bibfield  {journal}
  {\bibinfo  {journal} {JHEP}\ }\textbf {\bibinfo {volume} {08}}\bibfield
  {number} {\bibinfo  {number} { (20 08)},\ \bibinfo {pages} {007}},\ }\Eprint
  {https://arxiv.org/abs/0804.1773} {arXiv:0804.1773 [hep-th]} \BibitemShut
  {NoStop}%
\bibitem [{\citenamefont {Mukherjee}(2026)}]{MukherjeeWorkInProgress}%
  \BibitemOpen
  \bibfield  {author} {\bibinfo {author} {\bibfnamefont {J.}~\bibnamefont
  {Mukherjee}},\ }\href@noop {} {} (\bibinfo {year} {2026}),\ \bibinfo {note}
  {work in progress}\BibitemShut {NoStop}%
\bibitem [{\citenamefont {Schwimmer}\ and\ \citenamefont
  {Theisen}(2011)}]{SchwimmerTheisen}%
  \BibitemOpen
  \bibfield  {author} {\bibinfo {author} {\bibfnamefont {A.}~\bibnamefont
  {Schwimmer}}\ and\ \bibinfo {author} {\bibfnamefont {S.}~\bibnamefont
  {Theisen}},\ }\href {https://doi.org/10.1016/j.nuclphysb.2011.02.003}
  {\bibfield  {journal} {\bibinfo  {journal} {Nucl. Phys. B}\ }\textbf
  {\bibinfo {volume} {847}},\ \bibinfo {pages} {590} (\bibinfo {year}
  {2011})},\ \Eprint {https://arxiv.org/abs/1011.0696} {arXiv:1011.0696
  [hep-th]} \BibitemShut {NoStop}%
\end{thebibliography}%


\begin{thebibliography}{13}%
\makeatletter
\providecommand \@ifxundefined [1]{%
 \@ifx{#1\undefined}
}%
\providecommand \@ifnum [1]{%
 \ifnum #1\expandafter \@firstoftwo
 \else \expandafter \@secondoftwo
 \fi
}%
\providecommand \@ifx [1]{%
 \ifx #1\expandafter \@firstoftwo
 \else \expandafter \@secondoftwo
 \fi
}%
\providecommand \natexlab [1]{#1}%
\providecommand \enquote  [1]{``#1''}%
\providecommand \bibnamefont  [1]{#1}%
\providecommand \bibfnamefont [1]{#1}%
\providecommand \citenamefont [1]{#1}%
\providecommand \href@noop [0]{\@secondoftwo}%
\providecommand \href [0]{\begingroup \@sanitize@url \@href}%
\providecommand \@href[1]{\@@startlink{#1}\@@href}%
\providecommand \@@href[1]{\endgroup#1\@@endlink}%
\providecommand \@sanitize@url [0]{\catcode `\\12\catcode `\$12\catcode
  `\&12\catcode `\#12\catcode `\^12\catcode `\_12\catcode `\%12\relax}%
\providecommand \@@startlink[1]{}%
\providecommand \@@endlink[0]{}%
\providecommand \url  [0]{\begingroup\@sanitize@url \@url }%
\providecommand \@url [1]{\endgroup\@href {#1}{\urlprefix }}%
\providecommand \urlprefix  [0]{URL }%
\providecommand \Eprint [0]{\href }%
\providecommand \doibase [0]{https://doi.org/}%
\providecommand \selectlanguage [0]{\@gobble}%
\providecommand \bibinfo  [0]{\@secondoftwo}%
\providecommand \bibfield  [0]{\@secondoftwo}%
\providecommand \translation [1]{[#1]}%
\providecommand \BibitemOpen [0]{}%
\providecommand \bibitemStop [0]{}%
\providecommand \bibitemNoStop [0]{.\EOS\space}%
\providecommand \EOS [0]{\spacefactor3000\relax}%
\providecommand \BibitemShut  [1]{\csname bibitem#1\endcsname}%
\let\auto@bib@innerbib\@empty
\bibitem [{\citenamefont {Casini}\ \emph {et~al.}(2011)\citenamefont {Casini},
  \citenamefont {Huerta},\ and\ \citenamefont {Myers}}]{CHM}%
  \BibitemOpen
  \bibfield  {author} {\bibinfo {author} {\bibfnamefont {H.}~\bibnamefont
  {Casini}}, \bibinfo {author} {\bibfnamefont {M.}~\bibnamefont {Huerta}},\
  and\ \bibinfo {author} {\bibfnamefont {R.~C.}\ \bibnamefont {Myers}},\ }\href
  {https://doi.org/10.1007/JHEP05(2011)036} {\bibfield  {journal} {\bibinfo
  {journal} {JHEP}\ }\textbf {\bibinfo {volume} {05}}\bibfield  {number}
  {\bibinfo  {number} { (2011)},\ \bibinfo {pages} {036}},\ }\Eprint
  {https://arxiv.org/abs/1102.0440} {arXiv:1102.0440 [hep-th]} \BibitemShut
  {NoStop}%
\bibitem [{\citenamefont {Advant}\ \emph {et~al.}(2026)\citenamefont {Advant},
  \citenamefont {Anand}, \citenamefont {Benjamin}, \citenamefont {Kumar},
  \citenamefont {Minwalla}, \citenamefont {Mukherjee}, \citenamefont {Pal},
  \citenamefont {Rahaman},\ and\ \citenamefont {Ray}}]{FromZtoa}%
  \BibitemOpen
  \bibfield  {author} {\bibinfo {author} {\bibfnamefont {A.}~\bibnamefont
  {Advant}}, \bibinfo {author} {\bibfnamefont {H.}~\bibnamefont {Anand}},
  \bibinfo {author} {\bibfnamefont {N.}~\bibnamefont {Benjamin}}, \bibinfo
  {author} {\bibfnamefont {V.}~\bibnamefont {Kumar}}, \bibinfo {author}
  {\bibfnamefont {S.}~\bibnamefont {Minwalla}}, \bibinfo {author}
  {\bibfnamefont {J.}~\bibnamefont {Mukherjee}}, \bibinfo {author}
  {\bibfnamefont {S.}~\bibnamefont {Pal}}, \bibinfo {author} {\bibfnamefont
  {A.}~\bibnamefont {Rahaman}},\ and\ \bibinfo {author} {\bibfnamefont
  {P.}~\bibnamefont {Ray}},\ }\href@noop {} {\bibinfo {title} {From {$Z$} to
  {$a$}: High-temperature relations, subleading semi-universality, and
  conformal anomalies}} (\bibinfo {year} {2026}),\ \Eprint
  {https://arxiv.org/abs/2609.03013} {arXiv:2609.03013 [hep-th]} \BibitemShut
  {NoStop}%
\bibitem [{\citenamefont {Deser}\ and\ \citenamefont
  {Schwimmer}(1993)}]{DeserSchwimmer}%
  \BibitemOpen
  \bibfield  {author} {\bibinfo {author} {\bibfnamefont {S.}~\bibnamefont
  {Deser}}\ and\ \bibinfo {author} {\bibfnamefont {A.}~\bibnamefont
  {Schwimmer}},\ }\href {https://doi.org/10.1016/0370-2693(93)90934-A}
  {\bibfield  {journal} {\bibinfo  {journal} {Phys. Lett. B}\ }\textbf
  {\bibinfo {volume} {309}},\ \bibinfo {pages} {279} (\bibinfo {year}
  {1993})},\ \Eprint {https://arxiv.org/abs/hep-th/9302047}
  {arXiv:hep-th/9302047} \BibitemShut {NoStop}%
\bibitem [{\citenamefont {Schwimmer}\ and\ \citenamefont
  {Theisen}(2011)}]{SchwimmerTheisen}%
  \BibitemOpen
  \bibfield  {author} {\bibinfo {author} {\bibfnamefont {A.}~\bibnamefont
  {Schwimmer}}\ and\ \bibinfo {author} {\bibfnamefont {S.}~\bibnamefont
  {Theisen}},\ }\href {https://doi.org/10.1016/j.nuclphysb.2011.02.003}
  {\bibfield  {journal} {\bibinfo  {journal} {Nucl. Phys. B}\ }\textbf
  {\bibinfo {volume} {847}},\ \bibinfo {pages} {590} (\bibinfo {year}
  {2011})},\ \Eprint {https://arxiv.org/abs/1011.0696} {arXiv:1011.0696
  [hep-th]} \BibitemShut {NoStop}%
\bibitem [{\citenamefont {Skenderis}(2002)}]{Skenderis}%
  \BibitemOpen
  \bibfield  {author} {\bibinfo {author} {\bibfnamefont {K.}~\bibnamefont
  {Skenderis}},\ }\href {https://doi.org/10.1088/0264-9381/19/22/306}
  {\bibfield  {journal} {\bibinfo  {journal} {Class. Quant. Grav.}\ }\textbf
  {\bibinfo {volume} {19}},\ \bibinfo {pages} {5849} (\bibinfo {year}
  {2002})},\ \Eprint {https://arxiv.org/abs/hep-th/0209067}
  {arXiv:hep-th/0209067} \BibitemShut {NoStop}%
\bibitem [{\citenamefont {Dehghani}(2002)}]{Dehghani}%
  \BibitemOpen
  \bibfield  {author} {\bibinfo {author} {\bibfnamefont {M.~H.}\ \bibnamefont
  {Dehghani}},\ }\href {https://doi.org/10.1103/PhysRevD.65.124002} {\bibfield
  {journal} {\bibinfo  {journal} {Phys. Rev. D}\ }\textbf {\bibinfo {volume}
  {65}},\ \bibinfo {pages} {124002} (\bibinfo {year} {2002})},\ \Eprint
  {https://arxiv.org/abs/hep-th/0203049} {arXiv:hep-th/0203049} \BibitemShut
  {NoStop}%
\bibitem [{\citenamefont {He}(2022)}]{HeRotating}%
  \BibitemOpen
  \bibfield  {author} {\bibinfo {author} {\bibfnamefont {B.}~\bibnamefont
  {He}},\ }\href@noop {} {\bibinfo {title} {Holographic calculations of
  {R\'enyi} entropy from rotating topological black holes}} (\bibinfo {year}
  {2022}),\ \Eprint {https://arxiv.org/abs/2207.05354} {arXiv:2207.05354
  [hep-th]} \BibitemShut {NoStop}%
\bibitem [{\citenamefont {Balasubramanian}\ and\ \citenamefont
  {Kraus}(1999)}]{BalasubramanianKraus}%
  \BibitemOpen
  \bibfield  {author} {\bibinfo {author} {\bibfnamefont {V.}~\bibnamefont
  {Balasubramanian}}\ and\ \bibinfo {author} {\bibfnamefont {P.}~\bibnamefont
  {Kraus}},\ }\href {https://doi.org/10.1007/s002200050764} {\bibfield
  {journal} {\bibinfo  {journal} {Commun. Math. Phys.}\ }\textbf {\bibinfo
  {volume} {208}},\ \bibinfo {pages} {413} (\bibinfo {year} {1999})},\ \Eprint
  {https://arxiv.org/abs/hep-th/9902121} {arXiv:hep-th/9902121} \BibitemShut
  {NoStop}%
\bibitem [{\citenamefont {Henningson}\ and\ \citenamefont
  {Skenderis}(1998)}]{HenningsonSkenderis}%
  \BibitemOpen
  \bibfield  {author} {\bibinfo {author} {\bibfnamefont {M.}~\bibnamefont
  {Henningson}}\ and\ \bibinfo {author} {\bibfnamefont {K.}~\bibnamefont
  {Skenderis}},\ }\href {https://doi.org/10.1088/1126-6708/1998/07/023}
  {\bibfield  {journal} {\bibinfo  {journal} {JHEP}\ }\textbf {\bibinfo
  {volume} {07}}\bibfield  {number} {\bibinfo  {number} { (1998)},\ \bibinfo
  {pages} {023}},\ }\Eprint {https://arxiv.org/abs/hep-th/9806087}
  {arXiv:hep-th/9806087} \BibitemShut {NoStop}%
\bibitem [{\citenamefont {Gibbons}\ \emph {et~al.}(2005)\citenamefont
  {Gibbons}, \citenamefont {Perry},\ and\ \citenamefont
  {Pope}}]{GibbonsPerryPope}%
  \BibitemOpen
  \bibfield  {author} {\bibinfo {author} {\bibfnamefont {G.~W.}\ \bibnamefont
  {Gibbons}}, \bibinfo {author} {\bibfnamefont {M.~J.}\ \bibnamefont {Perry}},\
  and\ \bibinfo {author} {\bibfnamefont {C.~N.}\ \bibnamefont {Pope}},\ }\href
  {https://doi.org/10.1088/0264-9381/22/9/002} {\bibfield  {journal} {\bibinfo
  {journal} {Class. Quant. Grav.}\ }\textbf {\bibinfo {volume} {22}},\ \bibinfo
  {pages} {1503} (\bibinfo {year} {2005})},\ \Eprint
  {https://arxiv.org/abs/hep-th/0408217} {arXiv:hep-th/0408217} \BibitemShut
  {NoStop}%
\bibitem [{\citenamefont {Mukherjee}(2026)}]{MukherjeeWorkInProgress}%
  \BibitemOpen
  \bibfield  {author} {\bibinfo {author} {\bibfnamefont {J.}~\bibnamefont
  {Mukherjee}},\ }\href@noop {} {} (\bibinfo {year} {2026}),\ \bibinfo {note}
  {work in progress}\BibitemShut {NoStop}%
\bibitem [{\citenamefont {Giombi}\ \emph {et~al.}(2008)\citenamefont {Giombi},
  \citenamefont {Maloney},\ and\ \citenamefont {Yin}}]{GiombiMaloneyYin}%
  \BibitemOpen
  \bibfield  {author} {\bibinfo {author} {\bibfnamefont {S.}~\bibnamefont
  {Giombi}}, \bibinfo {author} {\bibfnamefont {A.}~\bibnamefont {Maloney}},\
  and\ \bibinfo {author} {\bibfnamefont {X.}~\bibnamefont {Yin}},\ }\href
  {https://doi.org/10.1088/1126-6708/2008/08/007} {\bibfield  {journal}
  {\bibinfo  {journal} {JHEP}\ }\textbf {\bibinfo {volume} {08}}\bibfield
  {number} {\bibinfo  {number} { (20 08)},\ \bibinfo {pages} {007}},\ }\Eprint
  {https://arxiv.org/abs/0804.1773} {arXiv:0804.1773 [hep-th]} \BibitemShut
  {NoStop}%
\bibitem [{\citenamefont {Benjamin}\ \emph {et~al.}(2024)\citenamefont
  {Benjamin}, \citenamefont {Lee}, \citenamefont {Pal}, \citenamefont
  {Simmons-Duffin},\ and\ \citenamefont {Xu}}]{AngularFractals}%
  \BibitemOpen
  \bibfield  {author} {\bibinfo {author} {\bibfnamefont {N.}~\bibnamefont
  {Benjamin}}, \bibinfo {author} {\bibfnamefont {J.}~\bibnamefont {Lee}},
  \bibinfo {author} {\bibfnamefont {S.}~\bibnamefont {Pal}}, \bibinfo {author}
  {\bibfnamefont {D.}~\bibnamefont {Simmons-Duffin}},\ and\ \bibinfo {author}
  {\bibfnamefont {Y.}~\bibnamefont {Xu}},\ }\href
  {https://doi.org/10.1007/JHEP11(2024)134} {\bibfield  {journal} {\bibinfo
  {journal} {JHEP}\ }\textbf {\bibinfo {volume} {11}}\bibfield  {number}
  {\bibinfo  {number} { (2024)},\ \bibinfo {pages} {134}},\ }\Eprint
  {https://arxiv.org/abs/2405.17562} {arXiv:2405.17562 [hep-th]} \BibitemShut
  {NoStop}%
\end{thebibliography}%
\end{document}